\documentclass[sigconf]{acmart}

\copyrightyear{2023}
\acmYear{2023}
\setcopyright{rightsretained}
\acmConference[SIGIR '23]{Proceedings of the 46th International ACM SIGIR Conference on Research and Development in Information Retrieval}{July 23--27, 2023}{Taipei, Taiwan}
\acmBooktitle{Proceedings of the 46th International ACM SIGIR Conference on Research and Development in Information Retrieval (SIGIR '23), July 23--27, 2023, Taipei, Taiwan}\acmDOI{10.1145/3539618.3591785}
\acmISBN{978-1-4503-9408-6/23/07}

\makeatletter
\gdef\@copyrightpermission{
  \begin{minipage}{0.3\columnwidth}
   \href{https://creativecommons.org/licenses/by/4.0/}{\includegraphics[width=0.90\textwidth]{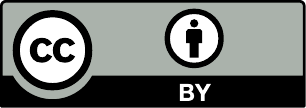}}
  \end{minipage}\hfill
  \begin{minipage}{0.7\columnwidth}
   \href{https://creativecommons.org/licenses/by/4.0/}{This work is licensed under a Creative Commons Attribution International 4.0 License.}
  \end{minipage}
  \vspace{5pt}
}
\makeatother

\usepackage{xcolor}

\usepackage{color, colortbl}
\usepackage{enumitem}

\usepackage{subcaption,booktabs}

\AtBeginDocument{}

\begin{document}
\title{When Newer is Not Better: Does Deep Learning Really Benefit Recommendation From Implicit Feedback?}

\vspace{-4ex}
\author{Yushun Dong}
\email{yd6eb@virginia.edu}
\affiliation{%
  \institution{University of Virginia}
  \country{}
}

\author{Jundong Li}
\email{jundong@virginia.edu}
\affiliation{%
  \institution{University of Virginia}
  \country{}
}

\author{Tobias Schnabel}
\email{toschnab@microsoft.com}
\affiliation{%
  \institution{Microsoft Research}
  \country{}
}

\renewcommand{\shortauthors}{Yushun Dong, Jundong Li, \& Tobias Schnabel}

\begin{abstract}

In recent years, neural models have been repeatedly touted to exhibit state-of-the-art performance in recommendation. Nevertheless, multiple recent studies have revealed that the reported state-of-the-art results of many neural recommendation models cannot be reliably replicated. A primary reason is that existing evaluations are performed under various inconsistent protocols. Correspondingly, these replicability issues make it difficult to understand how much benefit we can actually gain from these neural models. It then becomes clear that a fair and comprehensive performance comparison between traditional and neural models is needed.

Motivated by these issues, we perform a large-scale, systematic study to compare recent neural recommendation models against traditional ones in top-$n$ recommendation from implicit data. 
We propose a set of evaluation strategies for measuring memorization performance, generalization performance, and subgroup-specific performance of recommendation models.
We conduct extensive experiments with 13 popular recommendation models (including two neural models and 11 traditional ones as baselines) on nine commonly used datasets. Our experiments demonstrate that even with extensive hyper-parameter searches, neural models do not dominate traditional models in all aspects, e.g., they fare worse in terms of average HitRate. 
We further find that there are areas where neural models seem to outperform non-neural models, for example, in recommendation diversity and robustness between different subgroups of users and items.
Our work illuminates the relative advantages and disadvantages of neural models in recommendation and is therefore an important step towards building better recommender systems.

\end{abstract}



\maketitle

\section{Introduction}
Recent years have witnessed a surge of research interest in deep learning-based approaches for recommendation, including point-of-interest (POI) recommendation~\cite{yuan2013time,guo2018combining,sun2020go}, e-commerce recommendation~\cite{chen2019behavior,wang2018billion,schafer2001commerce}, and news recommendation~\cite{de2018news,wang2018dkn,zheng2018drn}, to name a few. Deep learning-based approaches parameterize user preferences through a series of neural network layers and then optimize ranking performance over an inventory of items~\cite{zhao2022revisiting,liang2018variational,he2017neural,he2020lightgcn}. 
With their large modeling capacity, neural models can, in theory, capture more complex and non-linear relationships from implicit user feedback. This, in turn, should increase the chances of finding a highly accurate recommendation model. 

Despite the undisputed success of neural models in many other areas, several studies have pointed out that the reported performance improvements of many neural recommendation models are difficult to reproduce~\cite {ferrari2019we,ludewig2019performance,ferrari2021troubling}. For example, several recent neural recommendation models are found to be outperformed by very simple recommendation models, such as linear models~\cite{ludewig2018evaluation,ferrari2019we} and those based on nearest neighbors~\cite{ludewig2018evaluation}. 
A primary reason for such replicability issues is that the reported performance advantages are concluded under many different experimental settings~\cite{zhao2022revisiting,cremonesi2021progress,zhao2021recbole,rendle2019difficulty}, such as dataset splitting ratios, evaluation protocols, metrics, recommendation tasks, etc.
This leaves us the important question of how much and what kind of benefits neural models can add to recommendation, which is a critical question for practitioners and system designers~\cite{da2020recommendation}. 
To answer this question, we need to have a more accurate understanding of when (i.e., under what scenarios) and how neural recommendation models and non-neural ones perform differently.

In this paper, we focus on exploring the performance differences between neural recommendation models and non-neural models in the context of top-$n$ item recommendation as one of the most common recommendation tasks~\cite{deshpande2004item}.
While there has been prior work on benchmarking top-$n$ recommender systems ~\cite{said2014comparative,cremonesi2010performance,adomavicius2016classification}, many of these efforts either did not include neural models or still left  important questions unanswered.
For example, recent work analyzing both neural and non-neural models~\cite{sun2022daisyrec,sun2020we} still lacks a thorough analysis of the performance difference between these two model classes and thus does not answer what benefits neural models may have over traditional ones.
Another line of work only investigates the performance differences between neural and non-neural recommendation models in limited aspects or settings~\cite{ferrari2019we,ludewig2019performance} and ignores other aspects where neural and non-neural models may differ. For instance, \cite{ferrari2019we} only focuses on the ranking accuracy over the test set. Ludwig et al. \cite{ludewig2019performance} compare neural and non-neural models under session-based recommendation tasks, ignoring the widely studied top-$n$ recommendation scenario.
More importantly, these benchmarks only focus on the \textit{weak generalization} setting~\cite{marlin2009collaborative,marlin2003modeling}, i.e., all users and items are known during training. This is clearly in contrast to real-world applications in which the recommender system  needs to accommodate newly joined users after deployment~\cite{jugovac2018streamingrec}. To move towards a more realistic scenario, we analyze performance in a strong generalization setting in this paper where new users are part of the test set.

To gain a deeper understanding of what benefits neural recommendation models can provide, this paper conducts a large-scale and systematic study of neural and non-neural models and explores a wide range of model properties beyond accuracy.
Specifically, we propose to empirically evaluate both the memorization and generalization abilities of recommendation models. By memorization, we mean the ability of a model to memorize the user data it has seen during training, while generalization refers to the ability to make accurate predictions  for new users during the test phase. As mentioned above, we adopt the setting of \textit{strong generalization}~\cite{marlin2003modeling,marlin2009collaborative} for generalization evaluation, where users are partitioned into non-overlapping sets for training, validation, and test, respectively. 
Note that there is usually a trade-off between memorization and generalization, which is also referred to as the bias-variance trade-off in machine learning~\cite{aggarwal2016recommender,wang2021denoising}. Robust recommendation performance usually requires a careful balance between them~\cite{cheng2016wide}.
We therefore join these two perspectives, and formally state our first research question as (\textbf{RQ1}) \textit{how do neural models differ from non-neural models in terms of both memorization and generalization?}
Moreover, most datasets are inhomogeneous and contain diverse subgroups of items and users with differing characteristics (e.g., item popularity and user preferences). Under such a context, a recommendation model that performs well on one subgroup may not perform well on others~\cite{ekstrand2018exploring,deldjoo2021explaining,ekstrand2012fairness}.
Hence an important dimension for evaluating the differences between neural and non-neural models is their recommendation performance over specific subgroups of users or items.
This understanding helps analyze the potential limitations of recommendation models, which is an important step in coming up with possible improvements in the future.
We state our second research question as (\textbf{RQ2}) \textit{how do neural models differ from non-neural models in terms of subgroup-specific performance?}

To answer the research questions above, we propose an array of practical and comprehensive evaluation strategies. 
We survey 13 popular recommendation models (including two neural models and 11 traditional models as baselines) and conduct extensive experiments over nine commonly used real-world datasets.
Experimental results indicate that even if hyper-parameters are sufficiently optimized, neural models do not outperform traditional models in terms of HitRate in most cases. 
However, we find that neural models can have certain advantages over traditional non-neural ones. For example, we find that neural models achieve better performance in terms of MeanRanks, recommendation diversity, map out semantic relationships between items more accurately, and show improved robustness between subgroups of instances (i.e., users and items).

To summarize, our contributions are three-fold: (1) \textbf{Experimental Design.} We propose a set of comprehensive evaluation strategies. To the best of our knowledge, our work serves as the first step towards evaluating top-$n$ recommendation performance differences between neural and non-neural models on both memorization and generalization. (2) \textbf{Performance Comparison.} We chose 13 popular recommendation models and nine commonly used datasets for our empirical investigation. We conduct a large-scale study assessing all dimensions defined above between neural recommendation models and non-neural ones. (3) \textbf{Comprehensive Analysis.} We present a comprehensive analysis of neural and non-neural recommendation models based on the experimental results. In addition, we also point out concrete directions for future research.

\section{Evaluation Strategies}

Through our experiments, we aim to answer the following two research questions in this paper:
\begin{description}
    \item[RQ1:] \textit{How do neural models differ from non-neural models in terms of performance on memorization and generalization tasks?}
\item[RQ2:] \textit{How do neural models differ from non-neural models in terms of subgroup-specific performance?}
\end{description}

Below we present notations and our strategies to evaluate memorization, generalization, and subgroup-specific performance.

\subsection{Notation}
\label{pre}

We use calligraphic letters (e.g., $\mathcal{A}$), bold lower-case letters (e.g., $\mathbf{a}$), and normal lower-case letters (e.g., $a$) to denote sets, vectors, and scalars, respectively.
In recommendation from implicit user feedback, the historical interactions between a specific user and item are binary, i.e., a user either likes an item or the relationship between them is unknown and can thus be expressed as a set.
We denote the set of users and items in a recommender system as $\mathcal{U} = \{U_1, ..., U_{|\mathcal{U}|}\}$ and $\mathcal{I} = \{I_1, ..., I_{|\mathcal{I}|}\}$, respectively.

In the memorization setting, we treat the whole set of users $\mathcal{U}$ as users visible during training (i.e., the training users), which is denoted as $\mathcal{U}_{\text{trn}}$.
In the strong generalization setting, we split the users into three non-overlapping sets, i.e., the user set for training $\mathcal{U}_{\text{trn}}$ ($\mathcal{U}_{\text{trn}} \subset \mathcal{U}$),  the user set for validation $\mathcal{U}_{\text{val}}$ ($\mathcal{U}_{\text{val}} \subset \mathcal{U}$), and the user set for test $\mathcal{U}_{\text{tst}}$ ($\mathcal{U}_{\text{tst}} \subset \mathcal{U}$, and $\mathcal{U}_{\text{trn}} \cup \mathcal{U}_{\text{val}} \cup \mathcal{U}_{\text{tst}}= \mathcal{U}$).

\subsection{Memorization Evaluation}
\label{memo}

Memorization loosely describes a model's ability to recall items seen during training~\cite{guo2017deepfm,feng2019deep,cheng2016wide}. Following this notion, we measure memorization in recommendation as a model's performance in recovering the implicit user feedback that was seen during training.
We now introduce two different tasks that let us probe a specific recommendation model for its memorization ability. Both tasks measure how well a model can recover an item that was seen during training but approach this goal in different ways.

\textbf{Leave-One-Out Memorization Task.}
In this task, we adapt the widely adopted leave-one-out protocol~\cite{li2011social,jamali2009trustwalker,jamali2009using,massa2007trust} to evaluate memorization. Different from the original protocol, we do not hold out a test item, but pick an item that was seen during training. Specifically, we randomly select one implicit feedback entry for each user in $\mathcal{U}_{\text{trn}}$. We then measure how close to the top a model ranks this entry when recommending items to users in $\mathcal{U}_{\text{trn}}$.

\textbf{Reranking Memorization Task.}
The setting of this task is similar to the one above, but here we focus on the performance of recovering \emph{all} training items for a user. More specifically, for each user in $\mathcal{U}_{\text{trn}}$, we measure whether  their training items will be ranked at the top when recommending items to this user.

To summarize, both tasks focus on performing evaluations over the set of training users $\mathcal{U}_{\text{trn}}$. The first strategy measures the ability of a recommendation model to predict single missing (hold-out) items, while the second strategy focuses on how well all training items can be recalled after fitting the model.

\subsection{Generalization Evaluation}
\label{generalization}

Generalization is loosely defined as applying the learned underlying data patterns to predict on unseen data~\cite{arpit2017closer,chatterjee2018learning,lecun2015deep,zhang2017understanding,cheng2016wide,koutrika2018modern}. 
Correspondingly, when we refer to the generalization performance of a recommendation model, we mean the performance of making accurate predictions on user-item combinations that were not seen during training.
It is worth noting that the generalization ability of a recommendation model has important real-world consequences as such models often have to make predictions for  new users~\cite{jugovac2018streamingrec}. 
We then discuss strategies to evaluate generalization below.

\textbf{Strong Generalization Task.}
We argue that strong generalization reflects the performance of recommendation models better in practice (compared with weak generalization). 
Specifically, evaluating the capability of strong generalization requires to test models on new users (i.e., those invisible ones during training) and existing items~\cite{marlin2003modeling}. 
Following the partitioning introduced in Section~\ref{pre}, we use users in $\mathcal{U}_{\text{trn}}$ with one hold-out item per user for training. After that, we use users in $\mathcal{U}_{\text{tst}}$ (with one hold-out item per user) for recommendation performance evaluation.

\textbf{Semantic Coherence under Strong Generalization.}
To gain a deeper understanding of the generalization performance, it is also critical to zoom in and examine how well those item-item semantics are captured~\cite{yao2018judging}.
Here, the intuition of our proposed strategy is that: given a user with only one interaction with a certain item, a model is considered to capture item-item semantics well if other items that are semantically similar (to this interacted item) can be identified and recommended.
Specifically, we propose to generate a set of dummy users $\tilde{\mathcal{U}}$. Each of these dummy users only interacts with one unique item out of $\mathcal{I}$ (thus $|\tilde{\mathcal{U}}| = |\mathcal{I}|$). For each dummy user $\tilde{U}_i \in \tilde{\mathcal{U}}$, we regard the generated user profile (i.e., the generated one user-item interaction) as the known historical implicit feedback, and employ the recommendation model to generate preference scores over all other items. 
In addition, we collect the semantics of all items (such as manual semantic tags) as side information.
For the item interacted with $\tilde{U}_i$, we also compute its similarity scores (e.g., based on cosine similarity) between the semantics of itself and that of all other items. 
Finally, we compute the Pearson correlation between the predicted preferences and the semantic similarity for $\tilde{U}_i$. 
We propose to consider the average Pearson correlation value over $\tilde{\mathcal{U}}$ as a general indicator of how well a recommendation model exploits the item-item semantics.

\begin{table*}[h]
\renewcommand{\arraystretch}{1.0}
\setlength\tabcolsep{6.5pt}
\centering
\caption{The statistics for the adopted datasets. Note that the presented statistics are the corresponding ones after pre-processing with binarizing and 5-core filtering. "Orig. Sparsity" indicates the level of sparsity before pre-processing.}
\label{statistics}
\begin{tabular}{cccccccc}
\hline
\textbf{Dataset} & \textbf{\#Users} & \textbf{\#Items} & \textbf{Orig. Sparsity} & \textbf{5-Core Sparsity} & \textbf{\#Ratings} & \textbf{\#Ratings Per Item} & \textbf{\#Ratings Per User} \\ \hline
\textbf{ml100k} &938 &1,008 &$6.30 \times 10^{-2}$  &$5.75 \times 10^{-2}$ &54.4 K &54.0 &58.0  \\ 
\textbf{lastfm} &1,859 &2,823 &$2.78 \times 10^{-3}$ &$1.36 \times 10^{-2}$ &71.4 K &25.3 &38.4  \\ 
\textbf{kuai} &1,411 &3,065 &$9.96 \times 10^{-1}$   &$5.01 \times 10^{-2}$ &216.7 K &70.7 &153.6  \\ 
\textbf{bookx} &13,854 &34,609 &$3.21 \times 10^{-5}$&$1.09 \times 10^{-3}$ &521.1 K &15.1 &37.6  \\ 
\textbf{ml1m} &6,034 &3,125 &$4.47 \times 10^{-2}$   &$3.05 \times 10^{-2}$ &574.4 K &183.8 &95.2  \\ 
\textbf{jester} &50,109 &100 &$5.63 \times 10^{-1}$ &$2.03 \times 10^{-1}$ &1.0 M &10,172.0 &20.3  \\ 
\textbf{amazon-e} &124,895 &44,843 &$3.91 \times 10^{-6}$ &$1.92 \times 10^{-4}$ &1.1 M &23.9 &8.6  \\ 
\textbf{ml20m} &136,674 &13,680 &$5.40 \times 10^{-3}$    &$5.34 \times 10^{-3}$ &10.0 M &729.3 &73.0  \\ 
\textbf{netflix} &463,435 &17,721 &$1.18 \times 10^{-2}$  &$6.93 \times 10^{-3}$ &56.9 M &3,209.7 &122.7  \\ \hline
\end{tabular}
\end{table*}

\renewcommand{\arraystretch}{1}

\subsection{Subgroup-Specific Performance Evaluation}
\label{subgroup}

A recommendation model may bear different performance over different subgroups of instances (e.g., users and items) for a specific dataset, where such subgroups are partitioned out of the original dataset w.r.t. certain instance-level characteristics. For example, when users are divided into active and inactive groups out of an e-commerce dataset, most recommendation models may deliver recommendations with significantly higher quality to the group of active users~\cite{rahmani2022unfairness,li2021user}.
Correspondingly, we define subgroup-specific performance as "\textit{the performance over subgroups of instances when the original dataset is partitioned following certain instance-level characteristics}".
In fact, one of the most widely studied partitionings is the warm-start subgroup versus cold-start subgroup~\cite{xu2012exploration,bu2016improving,abdullah2021eliciting,rana2014cold,hu2018leveraging}.
Motivated by this, we propose to partition warm-start and cold-start subgroups from the perspectives of both users and items.
In this section, we discuss our strategies to evaluate the subgroup-specific performance below.

\textbf{Partitioning Users: Active vs. Inactive.}
We propose to partition users into active user and inactive user subgroups according to their total number of interactions. We aim to explore the recommendation quality difference (across different models) over the two user subgroups. Specifically, we select the top $\frac{1}{3}$ active and top $\frac{1}{r}$ inactive users in $\mathcal{U}_{\text{tst}}$ to construct an active user set $\mathcal{U}_{\text{act}}$ and an inactive user set $\mathcal{U}_{\text{ina}}$, respectively. 
Here we propose to determine the value of $r$ by ensuring the total number of positive interactions in the two sets of users are generally the same, which ensures a fair comparison.
Given a specific recommendation model, we evaluate its recommendation performance on $\mathcal{U}_{\text{act}}$ and $\mathcal{U}_{\text{ina}}$ separately.

\textbf{Partitioning Users: Similar vs. Dissimilar.}
To evaluate the generalization capability at a finer granularity, we propose to divide $\mathcal{U}_{\text{tst}}$ into two groups based on the similarity of their implicit feedback with the users in $\mathcal{U}_{\text{trn}}$. Specifically, we first compute the similarity score between every test user (i.e., users in $\mathcal{U}_{\text{tst}}$) and training user (i.e., users in $\mathcal{U}_{\text{trn}}$). Then for each test user, we truncate its top-$l$ similarity scores with those training users and compute the sum of these truncated similarity scores. Here $l$ is a parameter that controls the computation of similarity truncation length.
We consider top $\frac{1}{3}$ test users with the largest values of the top-$l$ similarity score sum as similar users. In a similar vein, we consider top $\frac{1}{3}$ users with the smallest top-$l$ similarity scores sums as dissimilar users. Given a specific recommendation model, we evaluate its performance over the two sets of test users separately.

\textbf{Partitioning Items: Head vs. Tail.}
We finally explore how neural and non-neural recommendation models differ in recommendation quality for both head items (popular items with a large total number of interactions with test users) and tail items (items with fewer interactions).
Specifically, we first split the item set $\mathcal{I}$ into a head item set $\mathcal{I}_{\text{head}}$ and a tail item set $\mathcal{I}_{\text{tail}}$ according to the total number of users in $\mathcal{U}_{\text{tst}}$ who have interacted with them. The total interactions occupied by items from the two sets should generally be the same in order to ensure a fair comparison.
To perform evaluation, we propose to divide the test users into two sets determined by which set the items associated with their hold-out interactions belong to.
Given a specific recommendation model, we evaluate its performance over the two sets of test users separately.

\section{Benchmarked Models}
As previously mentioned, the generalization setting that this paper focuses on is strong generalization (as introduced in Section~\ref{generalization}). As a consequence, we only include the recommendation models that naturally support the strong generalization setting in our performance comparison.
Overall, we compare two neural models with five types of non-neural recommendation baselines: unpersonalized models, factorization-based models, nearest neighbor-based models, linear models, and graph-based models. 
\begin{description}[topsep=2mm]
\item[Neural Recommendation Models.] Only a few neural recommendation models are able to yield recommendations under the strong generalization setting. We selected MultiVAE and MultiDAE as they are both common as well as provide (near) state-of-the-art performance on some datasets~\cite{liang2018variational}. Both MultiVAE and MultiDAE are autoencoder-based recommendation models where the basic idea is to reconstruct the entire user profiles from partial versions. Extending MultiDAE, MultiVAE adopts a fully Bayesian approach to fit per-user variance, which could make it more prone to overfitting.

\item[Unpersonalized Models (Non-Neural).] Two unpersonalized recommendation models are adopted as baselines, namely Random and Popularity. Random generates a random permutation of items for each user, while Popularity generates recommendation results for users based on the popularity (total number of positive interactions) of items.

\item[Factorization-based Models (Non-Neural).] Two factorization-based models are selected, including PureSVD and ALS. Their main difference is that PureSVD performs a vanilla singular value decomposition on the user-item matrix, while ALS performs a weighted matrix factorization.

\item[Nearest Neighbor-based Models (Non-Neural).] ItemKNN and UserKNN are adopted as two traditional nearest neighbor-based recommendation models. Recommendations are derived based on the item-item and user-user similarity in the two models, respectively.

\item[Linear Models (Non-Neural).] SLIM~\cite{ning2011slim} and Ease~\cite{steck2019embarrassingly} are two popular linear baselines. Both of them aim to learn a linear function to capture the similarity for item-based collaborative filtering. However, SLIM aims to learn a sparse linear function, while Ease does not have the sparsity constraint. Hence Ease adopts the Frobenius norm of the learnable weight matrix as the regularization term instead of leveraging an $l_1$-norm as in SLIM.

\item[Graph-based Models (Non-Neural).] P3alpha~\cite{cooper2014random}, RP3beta~\cite{paudel2016updatable}, and Graph Filter based Collaborative Filtering (GFCF)~\cite{shen2021powerful} are adopted as the graph-based recommendation models. In general, these models consider the users and items as nodes, and the corresponding input matrix describes the existing edges between users and items with implicit user feedback.
Specifically, in P3alpha, items are ranked based on the reaching probability of a three-step walk to every user for recommendation; RP3beta is a modified version of P3alpha, where the outcomes are normalized by the corresponding item popularity; GFCF performs prediction based on the propagated input implicit user feedback matrix through existing edges.
\end{description}

\section{Experimental Setup}

\subsection{Datasets}

\textbf{Dataset Selection.}
A lot of previous work has published results only on a relatively small number of datasets~\cite{sun2020we}. Our guiding principle for dataset selection is to choose the most commonly used public datasets to ensure replicability.
Nine popular publicly available datasets are selected in this paper, namely ml100k~\cite{harper2015movielens}, ml1m~\cite{harper2015movielens}, ml20m~\cite{harper2015movielens}, lastfm~\cite{lastfm}, kuai~\cite{gao2022kuairec}, bookx~\cite{bookx}, jester~\cite{goldberg2001eigentaste}, amazon-e~\cite{mcauley2015image}, and netflix~\cite{bennett2007netflix}.

 \textbf{Pre-Processing Strategy.}
We conduct experiments based on implicit user feedback (such as binary entries representing clicks and purchases) and convert all ordinal rating data to binary. We follow the general consensus and treat all ratings greater or equal to four as positive and negative for the rest~\cite{sun2020we}.
In addition, it is also often necessary in practice to filter out users and items with insufficient ratings, i.e., to perform $h$-core filtering. We choose a common setting~\cite{zhao2022revisiting} to assign $h$ as five, and we present the statistics after filtering in Table~\ref{statistics}.

\subsection{Metrics}
\label{metrics}

We present our evaluation metric choices in this subsection. Specifically, we include three types of metrics, namely metrics for utility, diversity, and semantic coherence. We present the details below.

\textbf{Utility Metrics.}
There is a wide range of utility metrics that have been used to evaluate the ranking performance in top-$n$ recommendation. However, multiple popular metrics can be unstable during evaluation. For example, nDCG is affected by the total number of items and may flip the model comparison conclusion by only changing the gain factor used in computation~\cite{canamares2020offline}. Here, we adopt three commonly used metrics that are more stable to measure the recommendation utility, namely Recall@k, HitRate@k, and MeanRanks~\cite{fuhr2018some}. We define Recall@k for a user $U_i$ as
\begin{align}
    \text { Recall@k }= \frac{|R(U_i) \cap T(U_i)|}{|T(U_i)|}, \;\; U_i \in \mathcal{U},
\end{align}
where $R(U_i)$ denotes the set of retrieved items among the top-$n$ recommendations for user $U_i$; $T(U_i)$ represents the set of relevant items that are used during the test phase for user $U_i$. To be consistent with prior works, when $|T(U_i)| = 1$, we refer to Recall@k instead of HitRate@k. We present its formulation as
\begin{align}
    \text { HitRate@k } = |R(U_i) \cap T(U_i)|, \;\; U_i \in \mathcal{U}^*.
\end{align}
Here $\mathcal{U}^*$ is $\mathcal{U}$ and $\mathcal{U}_{\text{tst}}$ in the memorization and generalization setting, respectively.
Finally, we formulate MeanRanks as 
\begin{align}
    \operatorname{MeanRanks} = \frac{\sum_{I_i \in T(U_i)}\text{Rank}(I_i)}{|T(U_i)|}, \;\; U_i \in \mathcal{U}^*,
\end{align}
where the function $\text{Rank}(\cdot)$ takes an item as input and outputs the corresponding rank position in the output recommendations.
We leverage both Recall@k and MeanRanks to measure the model utility under the reranking memorization task. In other settings, we utilize HitRate@k and MeanRanks as the corresponding metrics.

\textbf{Diversity Metrics.}
Diversity is among the most popular beyond-accuracy ranking metrics~\cite{anelli2021elliot,valcarce2018robustness,zhang2008avoiding,qin2013promoting,eskandanian2020using,hurley2011novelty}. In our experiments, we employ the Gini Index~\cite{wilkie2014best} and Shannon Entropy~\cite{gray2011entropy} to measure diversity. The Gini Index comes from  economics, where it was originally used to measure disparity.
To compute it, one counts the number of times an item appears in the top-$n$ recommendations for users in the test set, and then normalizes this frequency so that the item frequencies form a probability distribution (e.g., $p(I_i)$ for item $I_i$). After sorting the items in increasing order $i_1, i_2, \ldots, i_n$, the Gini Index is defined as
\begin{align}
    \text{Gini\_Index}=\frac{1}{n-1} \sum_{j=1}^n(2 j-n-1) p\left(I_{i_j}\right).
\end{align}
As a second measure of diversity, we adopt Shannon Entropy. It is computed as follows: 
\begin{align}
\text{Shannon\_Entropy}=-\sum_{i=1}^n p(I_i) \log p(I_i).
\end{align}
We use both metrics above to measure diversity in order to get a more accurate picture of performance differences between models.

\textbf{Semantic Coherence Metrics.}
Another important aspect of differences between neural and non-neural models is the degree to which a model captures semantic relationships between items in the embedding space (see also  Section~\ref{generalization}). Inspired by user-centric studies such as~\cite{yao2018judging}, we propose a metric named Semantic Coherence Index (SCI). 
For each item $I_i \in \mathcal{I}$, we build a vector representation based on information that expresses human perception of similarity and relatedness (e.g., manual semantic tags).
Let $\mathbf{r}_i$ be the item similarity vector that a model produces. For models that use an item-item matrix, we simply use the corresponding entry. For models that do not have such a matrix, we input a one-hot user vector with entry $i$ set to one. We also compute the cosine similarity between the semantic vector of $I_i$ and that of every other item and denote the resulting vector as $\mathbf{s}_i$.
The SCI is then computed as
\begin{align}
\text{SCI} = \frac{1}{|\mathcal{I}|} \sum_{I_i \in \mathcal{I}} \text{Pearson} \left(\mathbf{s}_i, \mathbf{r}_i \right),
\end{align}
where the function $f$ is the recommendation model that takes two items as input, and outputs a vector depicting the predicted preference scores over all items in $\mathcal{I}$;
function $\text{Pearson}(\cdot)$ takes two vectors as input and outputs the value of their Pearson correlation.
We leverage SCI to measure how well recommendation models capture the item-item semantics.

\subsection{Experimental Settings}

\textbf{Evaluation Protocol.}
For each model, we perform training and testing under nested cross-validation with five user-based folds. For each round of evaluation, we use three folds as the training data, one fold for validation, and one fold for test.
We set the value for the truncation length $l$ for test-train user similarity ranking truncation (under subgroup-specific performance evaluation) to ten.

\textbf{Hyper-parameter Tuning.}
We conduct an extensive hyper-parameter search for all models. Specifically, the hyper-parameter search was done with Bayesian optimization for 50 iterations in every round of cross-validation. We used HitRate@50 as the target metric. 
The search space and the values of optimal hyper-parameters for each dataset in every round of cross-validation will be released upon acceptance.

\begin{figure*}[t]
\vspace{1mm}
        \begin{subfigure}[t]{0.9\textwidth}
        \centering
        \small
        \includegraphics[width=0.98\textwidth, height=0.035\textheight]{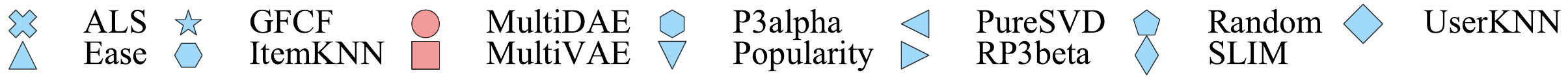}
        \end{subfigure}
\begin{subfigure}[t]{0.23\textwidth}
\small
\includegraphics[width=0.95\textwidth]{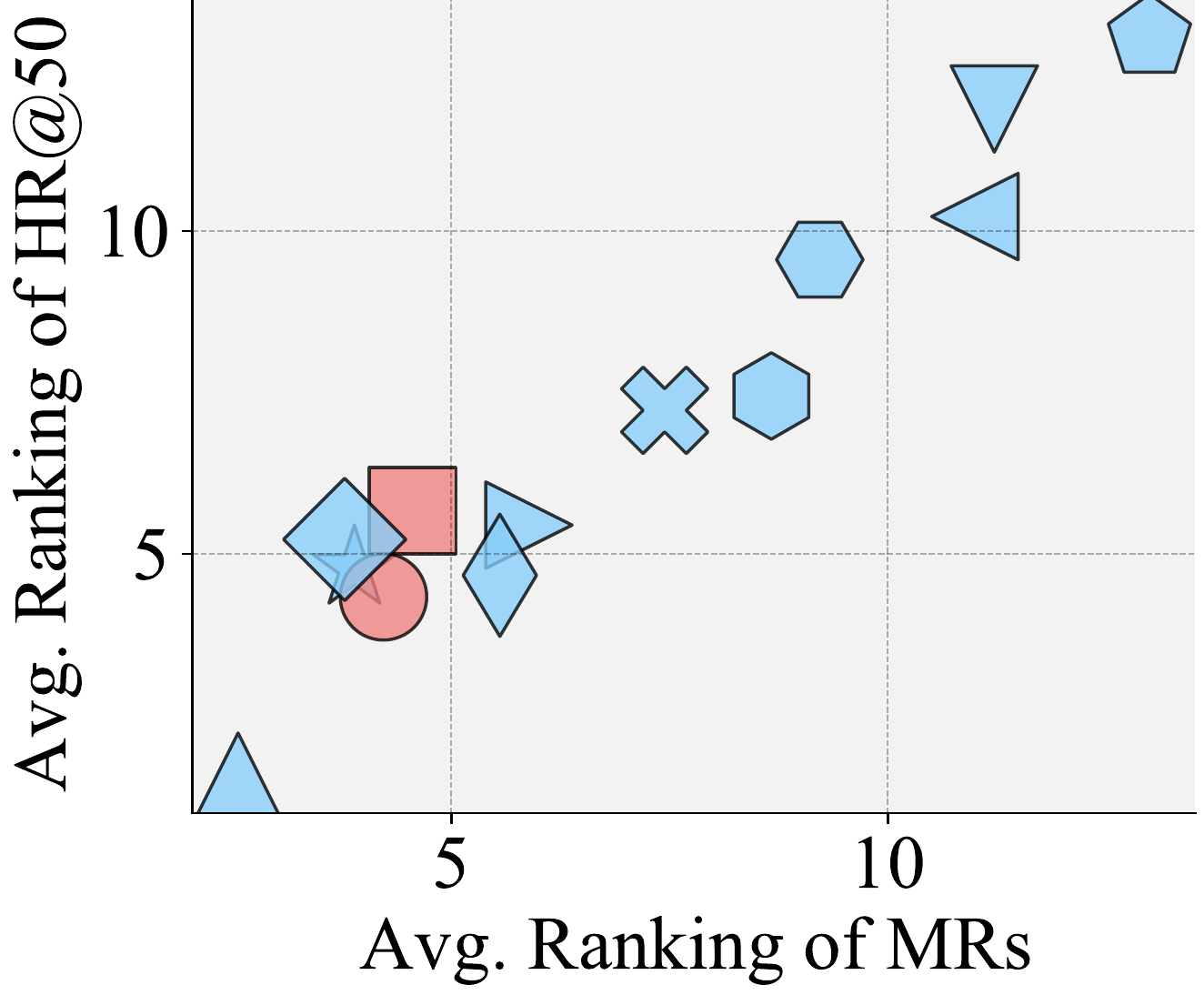}
\vspace{-2mm}
    \caption[Network2]%
    {{\footnotesize Leave-one-out memorization task}}    
    \label{memo1}
\end{subfigure} 
        \begin{subfigure}[t]{0.23\textwidth}
\small
\includegraphics[width=0.995\textwidth]{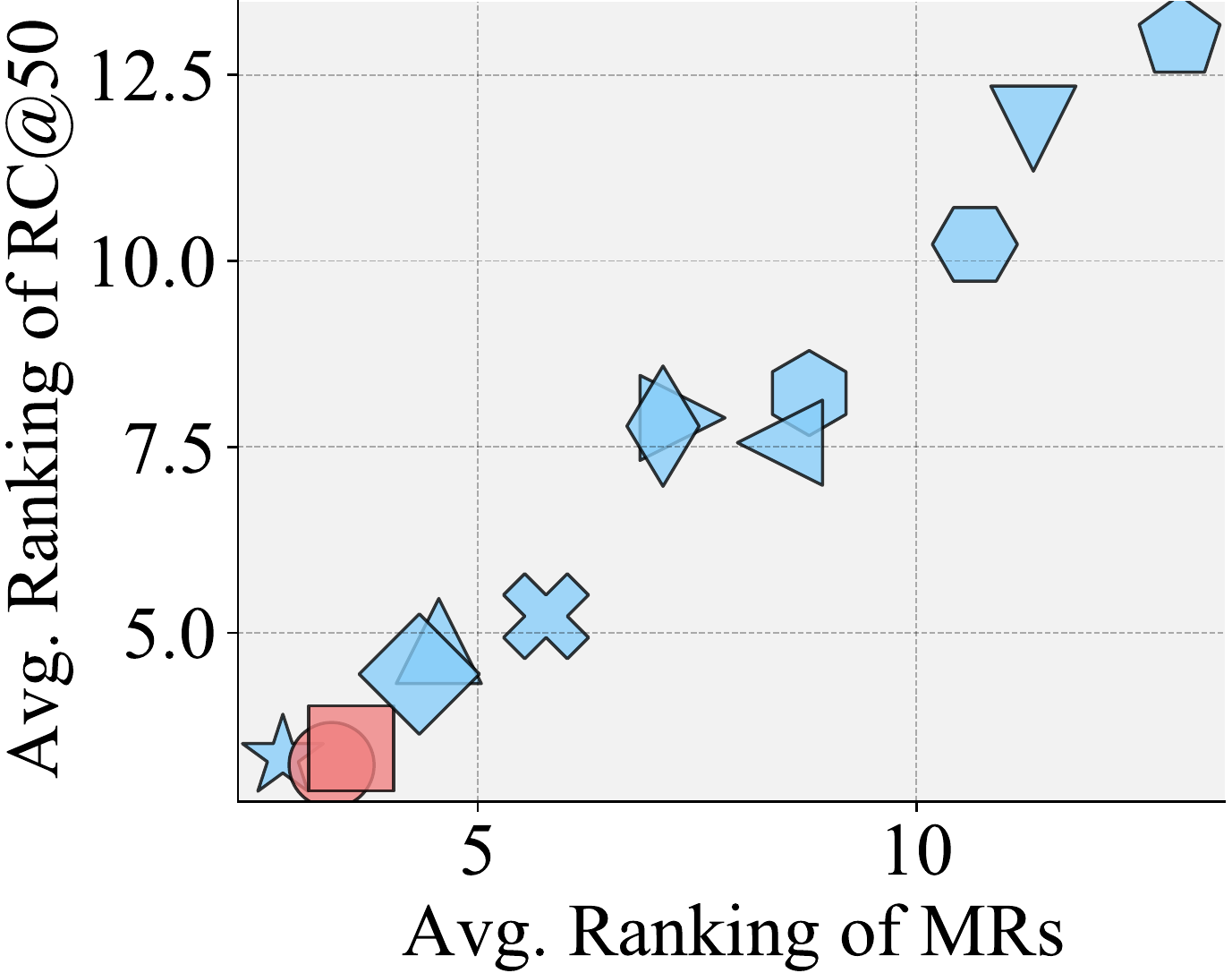}
\vspace{-5.5mm}
    \caption[Network2]%
    {{\footnotesize Reranking memorization task}}    
    \label{memo2}
\end{subfigure}
        \begin{subfigure}[t]{0.23\textwidth}
\small
\includegraphics[width=0.995\textwidth]{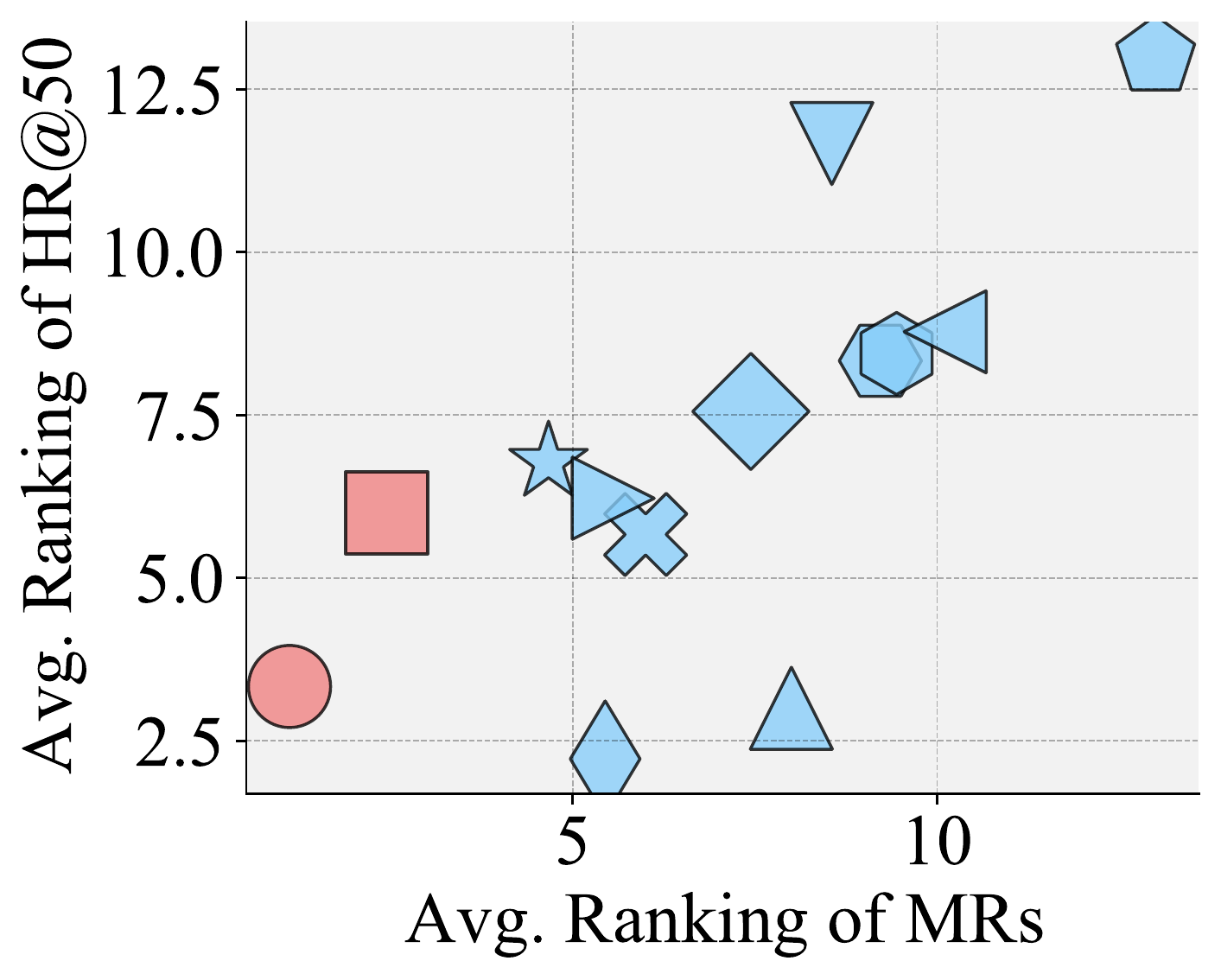}
\vspace{-5.5mm}
    \caption[Network2]%
    {{\footnotesize Strong generalization task}} 
    \label{gen}
\end{subfigure}
\begin{subfigure}[t]{0.225\textwidth}
\small
\includegraphics[width=0.995\textwidth]{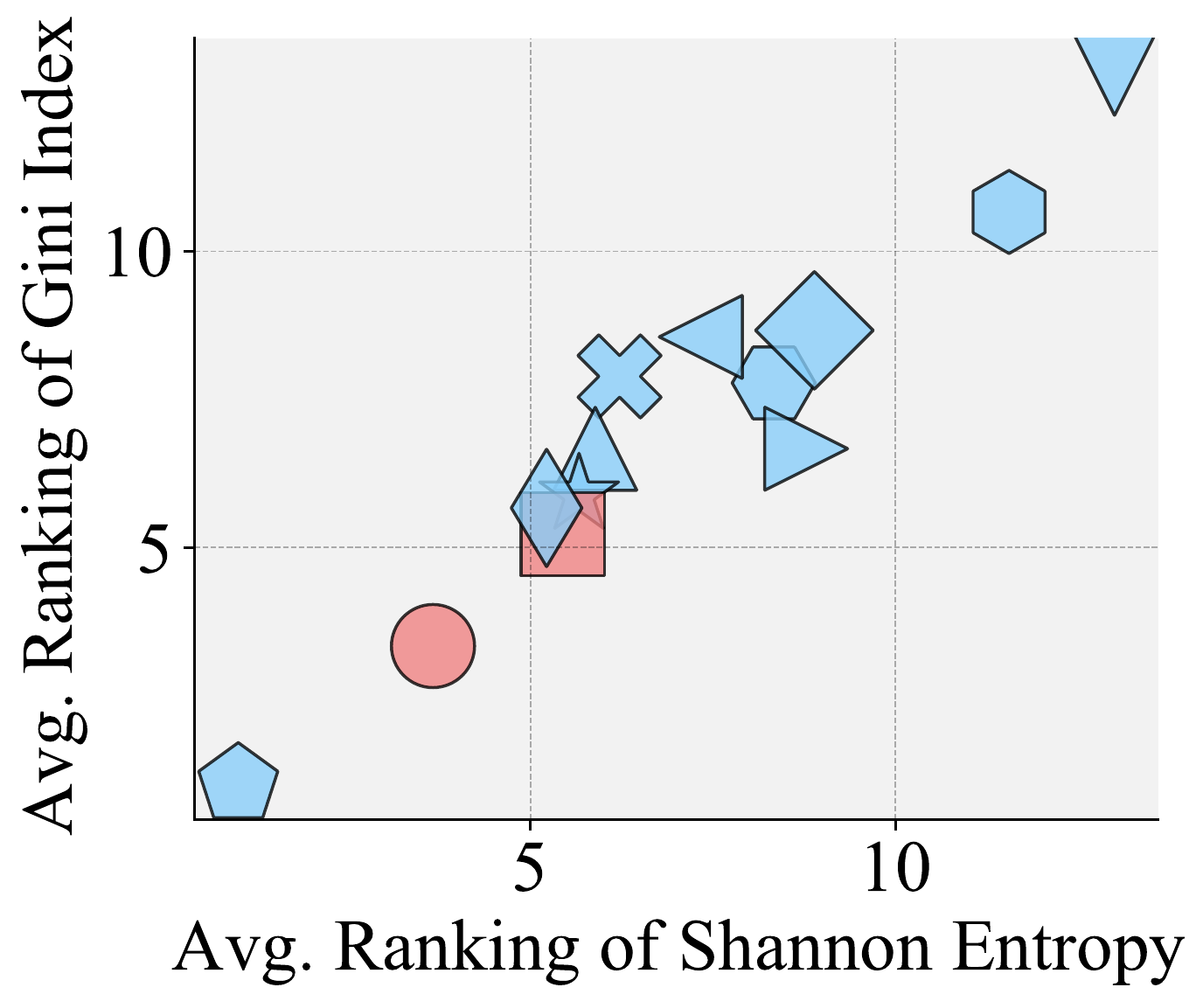}
\vspace{-5.5mm}
    \caption[Network2]%
    {{\footnotesize Recommendation diversity}}    
    \label{div}
\end{subfigure} 
        \vspace{-3mm}
\caption{In (a) and (b), we present the results of the two different memorization tasks, respectively. In (c), we show the generalization results in terms of recommendation utility. In (d), we present the results for recommendation diversity in terms of Gini Index and Shannon Entropy. We use "RC", "HR", and "MRs" to represent Recall, HitRate, and MeanRanks, respectively. Average rankings are computed over all nine datasets for every recommendation model. The two neural recommendation models are in light coral while others use a light blue.}
\label{memo_gen}
    \vspace{-2.5mm}
\end{figure*}

\begin{figure}[t]
\begin{subfigure}[t]{0.49\textwidth}
\small
\centering
\includegraphics[width=0.7\textwidth, height=0.0165\textheight]{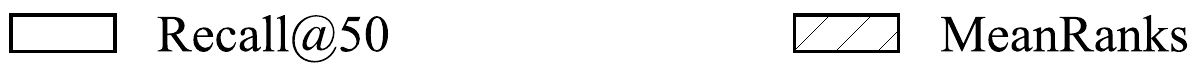}
\end{subfigure} 
\begin{subfigure}[t]{0.233\textwidth}
\small
\includegraphics[width=0.95\textwidth]{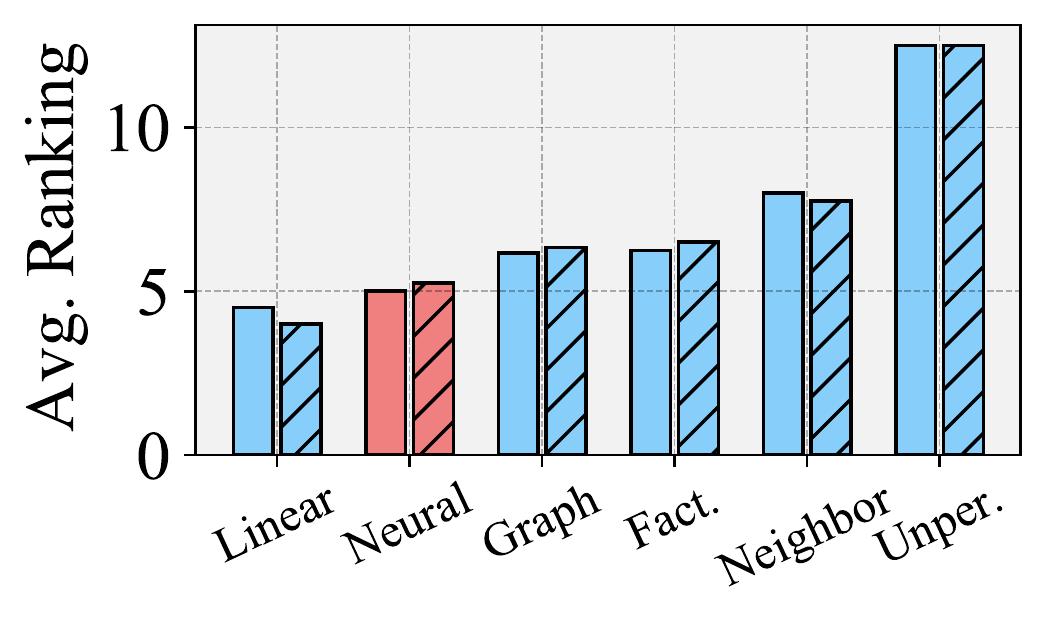}
        \vspace{-2mm}
    \caption[Network2]%
    {{\footnotesize Average ranking on the datasets  with top-two smallest sizes}}    
    \label{memo_small}
\end{subfigure} 
\hspace{1mm}
        \begin{subfigure}[t]{0.227\textwidth}
\small
\includegraphics[width=0.995\textwidth]{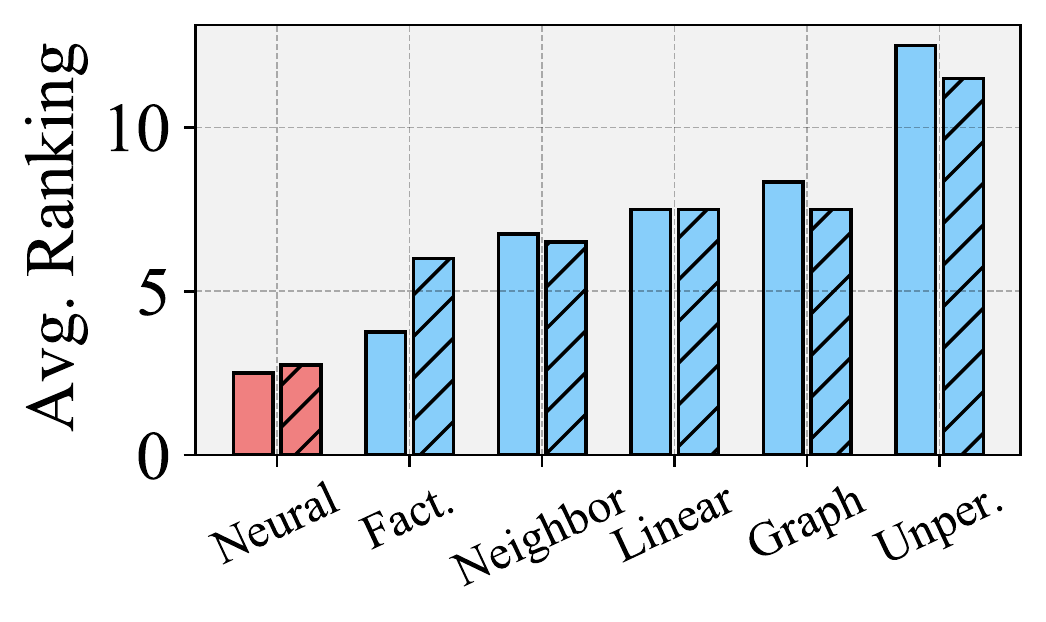}
        \vspace{-5.7mm}
    \caption[Network2]%
    {{\footnotesize Average ranking on the datasets with top-two largest sizes}} 
    \label{memo_large}
\end{subfigure}
        \vspace{-4mm}
\caption{Results of the reranking memorization task between neural models and other five groups of baselines. The performance of each baseline group is averaged over the top two smallest/largest datasets. The two neural models are in light coral and others are in light blue.}
\label{memo_small_large}
    \vspace{1mm}
\end{figure}

\begin{figure}[t]
\centering
\vspace{-4mm}
\includegraphics[width=0.45\textwidth]{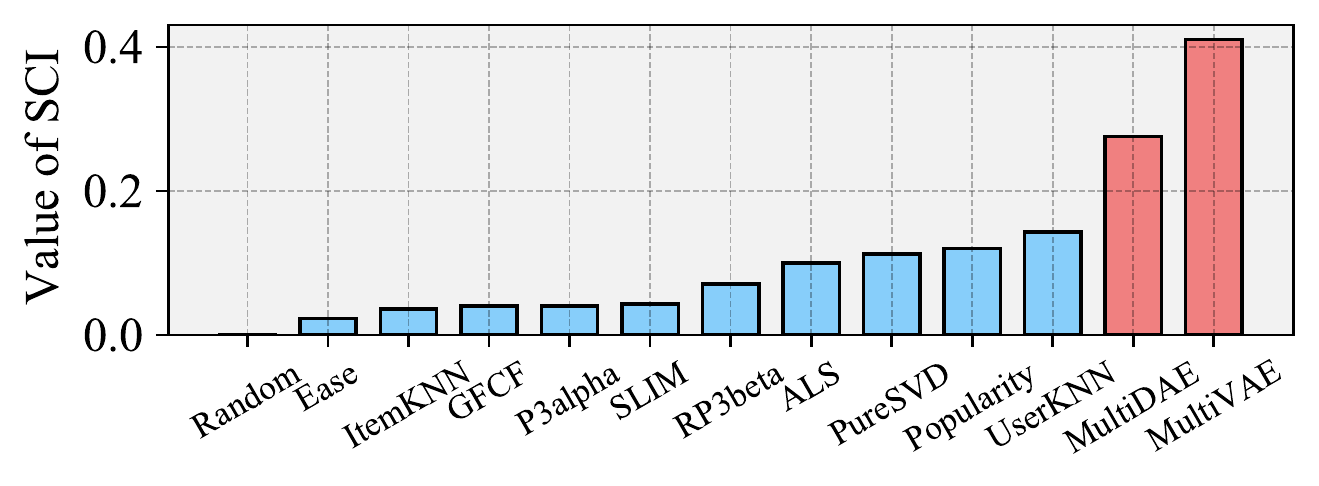}
        \vspace{-4mm}
\caption{Results for semantic coherence on the ml20m dataset. The two neural recommendation models are marked in light coral and others in light blue.}
\label{semantics}
    \vspace{-4mm}
\end{figure}

\section{Empirical Investigation}

\subsection{Finding 1: Neural models excel on datasets with larger sizes in memorization}

We first perform experiments on the two memorization settings introduced in Section~\ref{memo}. In the leave-one-out memorization task, we collect the values of HitRate@50 and Meanranks. In the reranking memorization task, we collect the values of Recall@50 and MeanRanks. We compute the ranking of all models for each dataset, and then rank them by their average rank over all datasets. We present the results in Fig.~\ref{memo_gen}(\subref{memo1}) and Fig.~\ref{memo_gen}(\subref{memo2}). 

 \textbf{Leave-One-Out Memorization Task.}
We first discuss the differences between neural and non-neural models in the leave-one-out memorization tasks. From the experimental results, we can observe that neural models do not exhibit superiority over other non-neural models, and simpler non-neural models are still able to outperform neural ones. For example, Ease, as a linear recommendation model, outperforms neural models on both utility metrics.

 \textbf{Reranking Memorization Task.}
We examine the differences between neural and non-neural models in the so-called reranking memorization task. We observe that neural models are among the ones with the best reranking performance.
A potential reason could be that the high model capacity enables neural models to perform well in memorizing all implicit user feedback that has been seen during training. 
Nevertheless, it is also worth noting that several non-neural models also perform well, e.g., linear models, which is consistent with the findings from existing work (e.g.,~\cite{ferrari2019we}) in weak generalization settings.

 \textbf{Memorization under Different Dataset Sizes.}
Based on the previous results, we further zoom in on the differences regarding the fitting capabilities of neural models and non-neural models. Specifically, we compute the average performance ranking in the reranking memorization task for each baseline model over datasets with the top-$k$ smallest and largest sizes (in terms of the total number of implicit user feedback).
We present the results for $k=2$ in Fig.~\ref{memo_small_large}(\subref{memo_small}) and Fig.~\ref{memo_small_large}(\subref{memo_large}).
We observe that linear recommendation models achieve the best performance over all other models on the top-$k$ smallest datasets.
Nevertheless, neural recommendation models achieve the best performance on the top-$k$ largest datasets.
Although not shown, these trends also hold for other values of $k$. One hypothesis for these results is that smaller datasets with less implicit user feedback (i.e., user-item interactions) tend to possess simpler user preference patterns. As a consequence, simpler models, such as linear ones, are able to memorize the user profiles well. Another explanation might be that as neural models are higher capacity, they need more data in order to not overfit.

To summarize the observations, we found that neural models did not show superior performance in our leave-one-out memorization task but were among the best-performing ones in the reranking memorization task. We hypothesize that neural models perform better in the reranking task because it emphasizes the performance on recovering all training items correctly at the same time. With their smooth parameterization, neural models may be able to do this soft clustering better than sparse linear models.

\subsection{Finding 2: Neural and non-neural models generalize differently}
\label{gen_eval}

We now turn to investigate how neural and non-neural models differ in their generalization abilities. Going beyond pure accuracy, we discuss their differences from three perspectives, including recommendation utility, diversity, and semantic coherence.

 \textbf{Recommendation Utility.}
This assesses recommendation utility by measuring HitRate and MeanRanks in the strong generalization setting. We first present a general summary of the rankings of each recommendation model in Fig.~\ref{memo_gen}(\subref{gen}). We observe that in terms of HitRate@50 (the vertical axis), although neural models are among the ones with the best performance, both linear models (Ease and SLIM) achieve better performance compared with the neural ones. Put another way: neural models did not perform  any better than traditional linear models.
However, when considering how far relevant items appear in the ranking by measuring MeanRanks (the horizontal axis), neural models are superior to all other recommendation models. These experimental results point again to the ability of neural models to have soft and coherent clusters -- similar to the reranking memorization task. 

In addition, we also present the detailed performance statistics in Table~\ref{all_gen_results}. These detailed results are with our previous summary: neural models tend to outperform other non-neural models in terms of MeanRanks, but they do not exhibit any superiority compared with linear models in terms of HitRate.

 \textbf{Recommendation Diversity.}
To study the performance differences in recommendation diversity, we computed the Gini Index and Shannon Entropy of the resulting recommendations. We look at the top-10 ($k = 10$) performance scores for more discriminative power and present the diversity rankings of each model in Fig.~\ref{memo_gen}(\subref{div}). Our observations are as follows. First and unsurprisingly, the unpersonalized Random method  achieves the best performance in terms of both metrics for recommendation diversity. However, Random is not an effective recommendation approach in real-world applications due to its poor utility, e.g., low HitRate and MeanRanks indicated in Fig.~\ref{memo_gen}(\subref{gen}). Among the remaining models, neural models achieve the best performance  on both diversity metrics. These results indicate that neural models deliver recommendations with a higher level of diversity, which implies that neural models are less prone to putting high scores on popular items.

 \textbf{Recommendation Semantic Coherence.}
We finally analyze the semantic coherence of recommendation, where we use side information about the items collected from user-specified tags. The results we present here are based on the MovieLens ml20m dataset, where human-annotated semantic information is available from the semantic tags submitted by users. We follow the methodology outlined in Section~\ref{generalization} and present the experimental results in Fig.~\ref{semantics}. We observe that neural models achieve the highest values in terms of Semantic Coherence Index. Such an observation suggests that neural models are better at capturing item-item semantics as they are learning from user feedback.

To summarize, we found that in the strong generalization setting, non-neural models (on average) achieve better HitRate, while neural models achieve better performance in terms of MeanRanks. This at least gives some guidance for practitioners -- non-neural models should be preferred when a better HitRate is desired.
In addition, neural models achieve a higher level of diversity and can better capture the item-item semantic relationship compared with non-neural ones. Thus neural models should be preferred when diversity and having adequate item-item relationships are of importance.

\begin{figure}[!t]
\begin{subfigure}[t]{0.45\textwidth}
\small
\flushright
\vspace{2mm}
\includegraphics[width=1.0\textwidth, height=0.044\textheight]{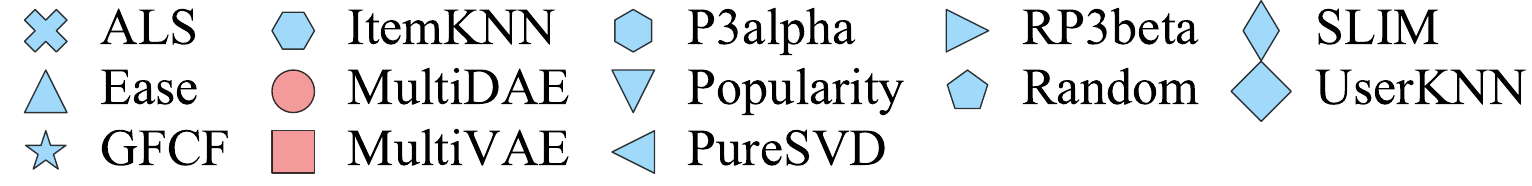}
\vspace{0.2mm}
\end{subfigure} 
        \begin{subfigure}[t]{0.23\textwidth}
        \small
        \includegraphics[width=0.995\textwidth]{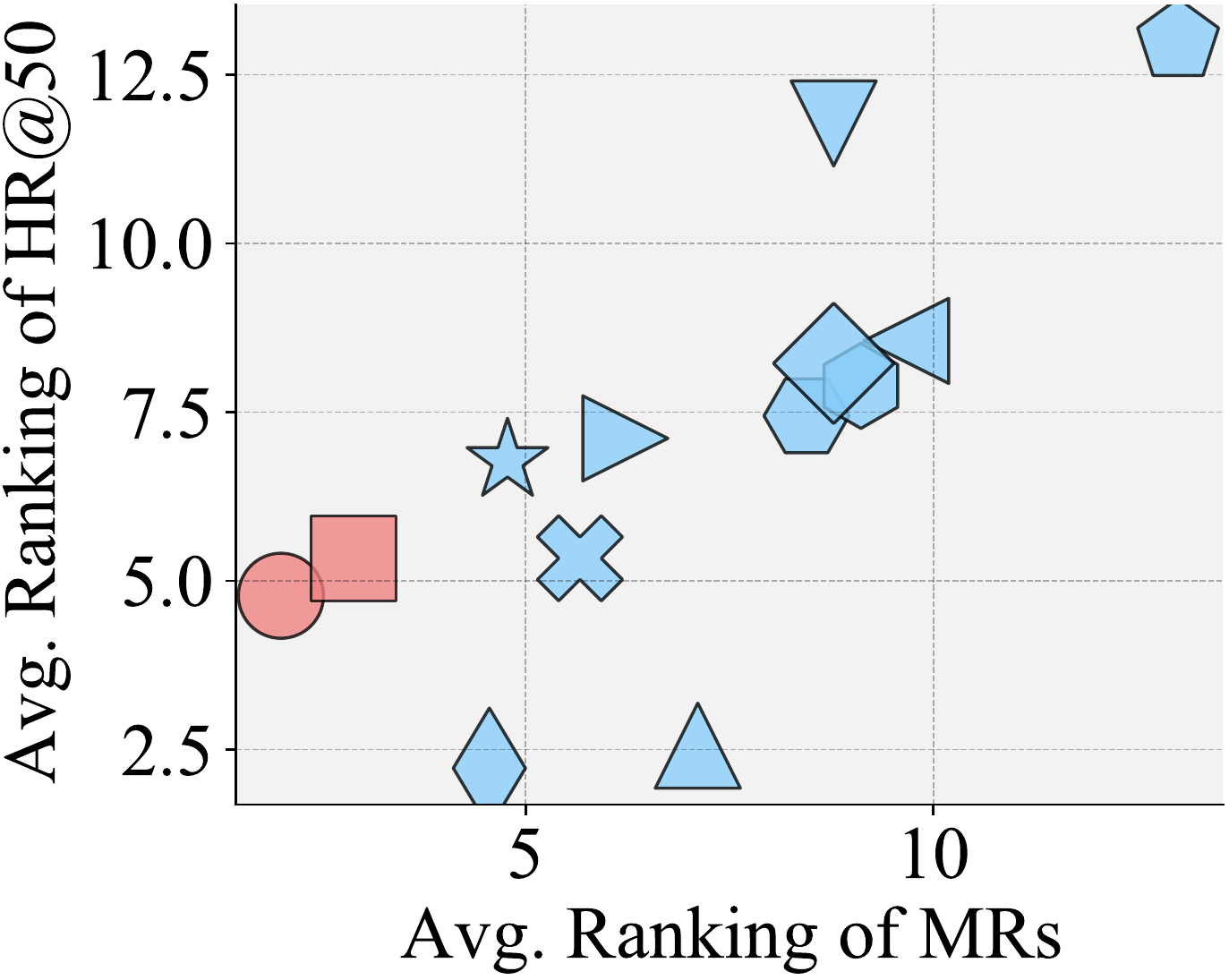}
        \vspace{-5mm}
            \caption[Network2]%
            {{\footnotesize \;\;\; Similar users}} 
            \label{easy}
        \end{subfigure}
        \begin{subfigure}[t]{0.23\textwidth}
        \small
        \includegraphics[width=0.995\textwidth]{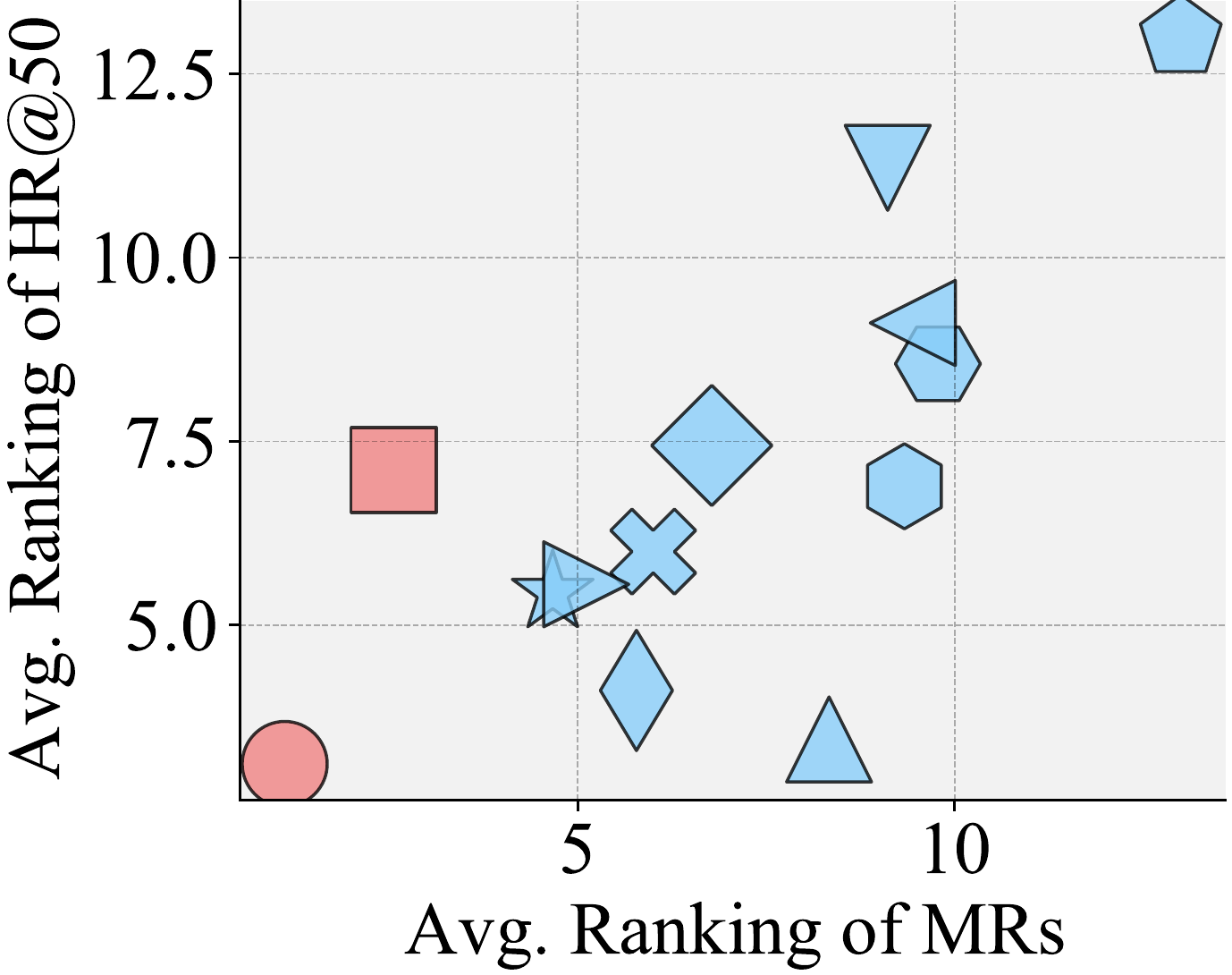}
        \vspace{-5mm}
            \caption[Network2]%
            {{\footnotesize \;\;\; Dissimilar users}} 
            \label{hard}
        \end{subfigure} \\
                \vspace{3mm}
        \begin{subfigure}[t]{0.23\textwidth}
        \small
        \includegraphics[width=0.995\textwidth]{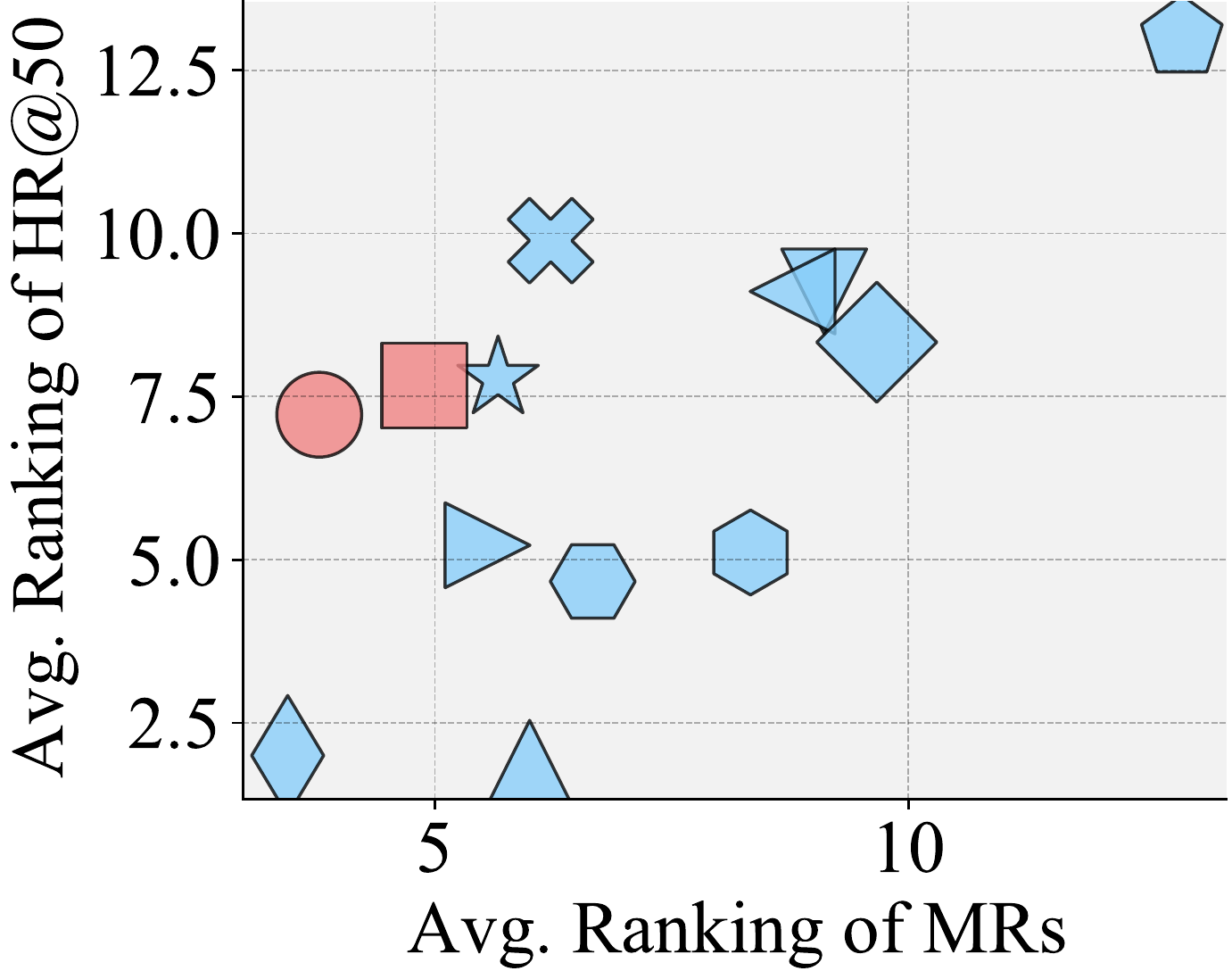}
        \vspace{-5mm}
            \caption[Network2]%
            {{\footnotesize \;\;\; Active users}}    
            \label{active}
        \end{subfigure} 
    \begin{subfigure}[t]{0.23\textwidth}
        \small
        \includegraphics[width=0.995\textwidth]{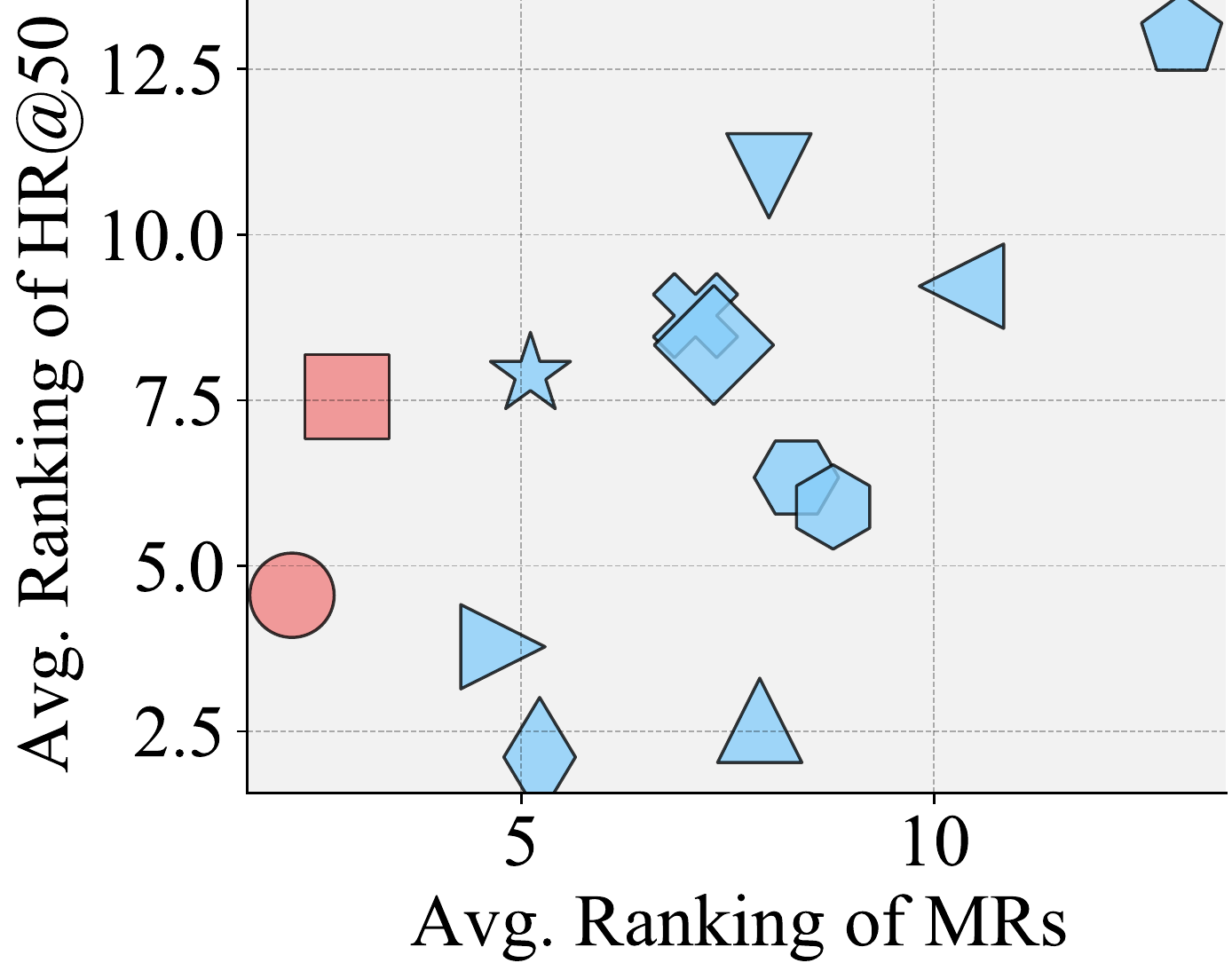}
        \vspace{-5mm}
            \caption[Network2]%
            {{\footnotesize \;\;\; Inactive users}} 
            \label{inactive}
        \end{subfigure} \\
                \vspace{3mm}
            \begin{subfigure}[t]{0.23\textwidth}
        \small
        \includegraphics[width=0.995\textwidth]{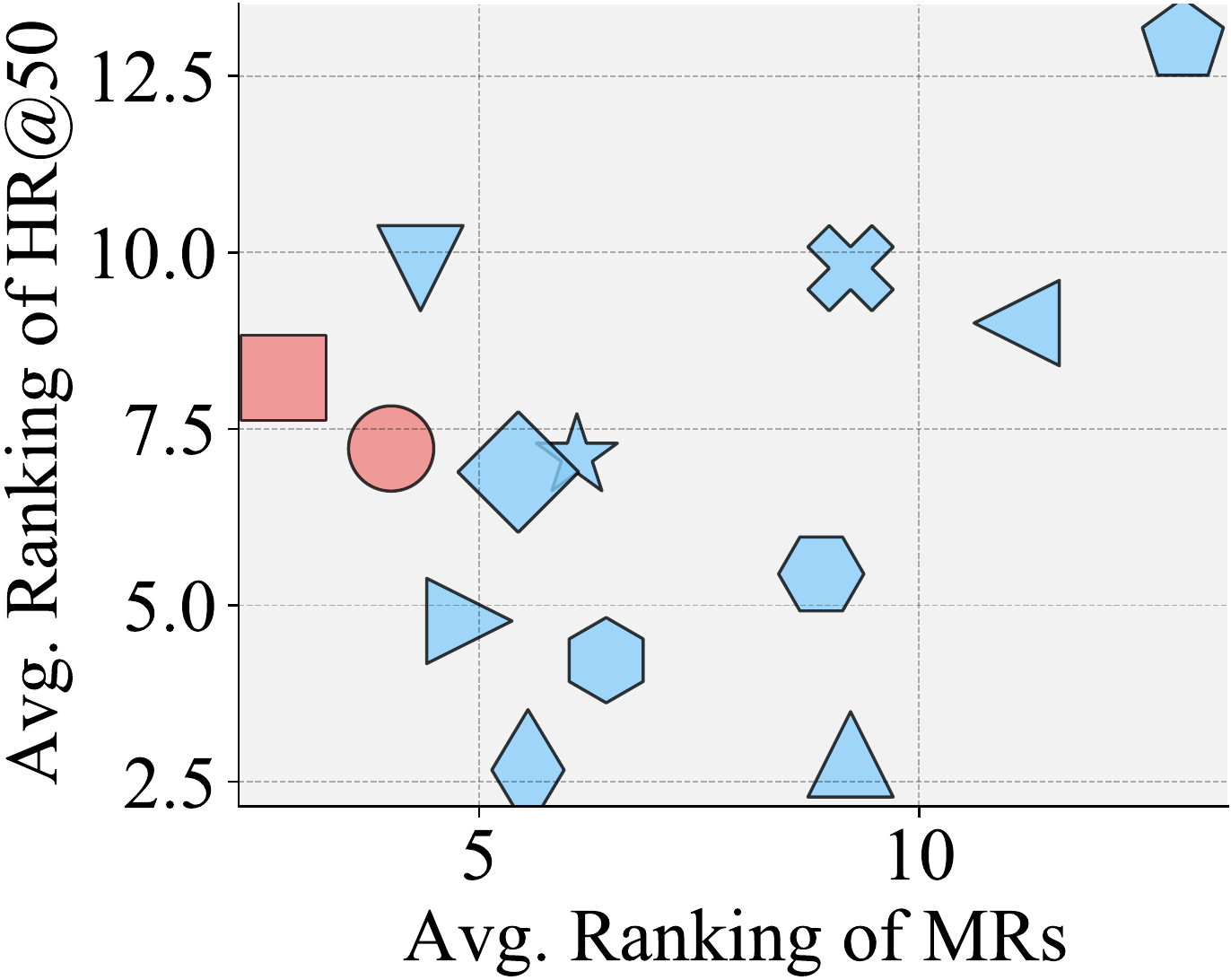}
        \vspace{-5mm}
            \caption[Network2]%
            {{\footnotesize \;\;\; Head items}}   
            \label{head}
        \end{subfigure} 
        \begin{subfigure}[t]{0.23\textwidth}
        \small
        \includegraphics[width=0.995\textwidth]{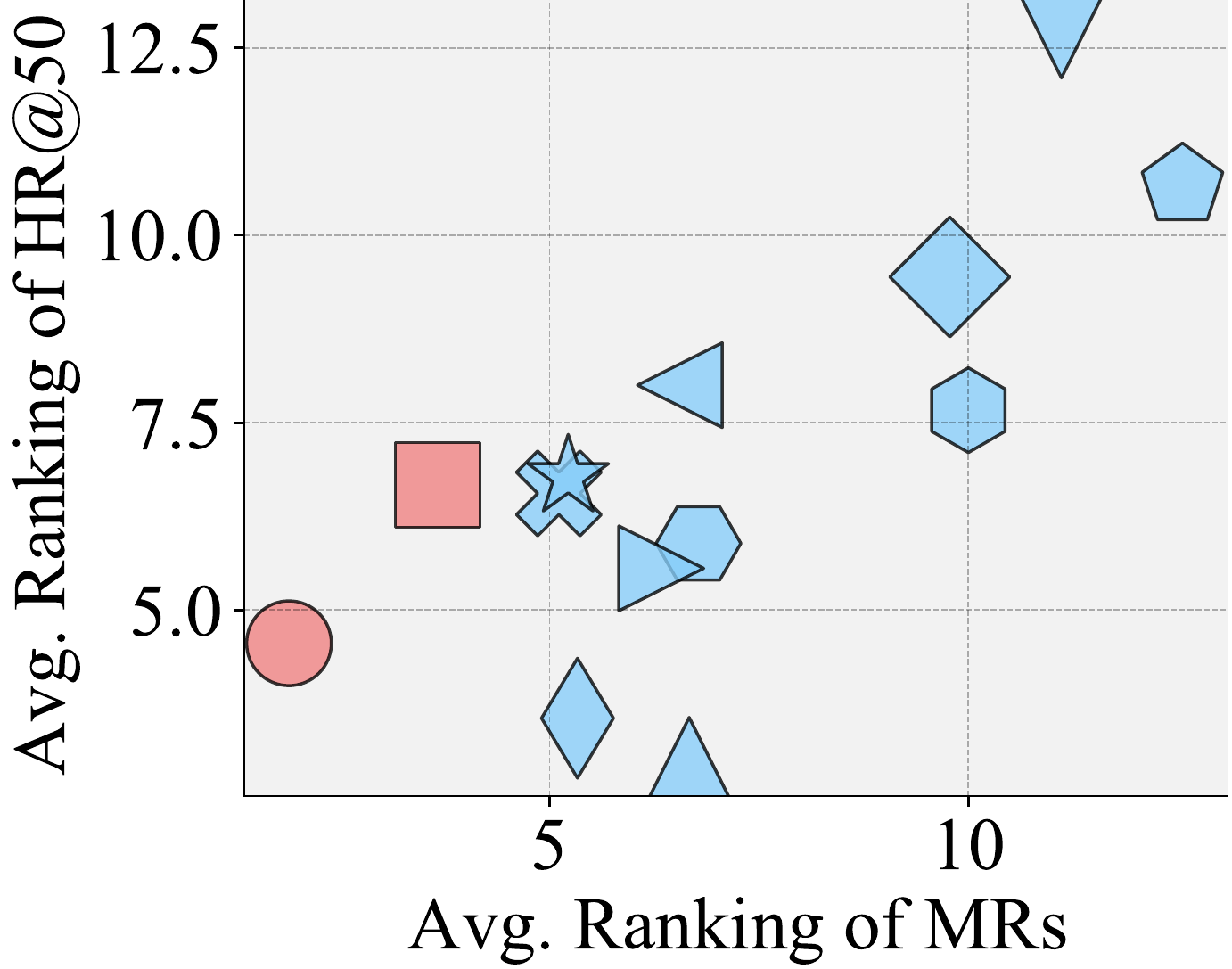}
        \vspace{-5mm}
            \caption[Network2]%
            {{\footnotesize \;\;\; Tail items}} 
            \label{tail}
        \end{subfigure}
        \vspace{-2mm}
    \caption{Evaluation of subgroup-specific performance. We use "HR" and "MRs" to represent HitRate and MeanRanks, respectively. The two neural models are marked out with color light coral against others as light blue.}
    \vspace{-4mm}
    \label{subgroup}
\end{figure}

\subsection{Finding 3: Neural models exhibit stronger robustness among different subgroups}

We now explore subgroup-specific performance differences between neural and non-neural models as discussed in Section~\ref{subgroup}. We examine performance differences between user subgroups and item subgroups in strong generalization settings as before.

 \textbf{Similar Users vs. Dissimilar Users.}
We first study the performance differences between similar and dissimilar users. Here similar users refer to test users that have the largest top-ranked cosine similarities with the users that have been seen during training (in terms of implicit user feedback), while dissimilar users are those who are most dissimilar to those training users. We present the experimental results over the two subgroups of users in Fig.~\ref{subgroup}(\subref{easy}) and Fig.~\ref{subgroup}(\subref{hard}), respectively. We make the following observations. First, neural and non-neural models mimic the performance ranking we found in the generalization experiments (introduced in Section~\ref{gen_eval}): neural models do better in terms of  MeanRanks but cannot outperform traditional linear recommendation models on HitRate. Second, relative to its performance over similar users, MultiDAE achieves better performance over dissimilar ones. Other models don't experience much change in terms of HitRate. However, in terms of MeanRanks, both neural models fare substantially better than non-neural models on dissimilar users. 
These findings indicate that neural models are more robust with respect to their performance across the two different user subgroups (i.e., similar and dissimilar), both for HitRate as well as MeanRanks.

\begin{table*}[!tbp]
\small
\renewcommand{\arraystretch}{0.8}
\setlength\tabcolsep{2.2pt}
  \centering
  \begin{subtable}[t]{0.32\textwidth}
    \begin{center}
    \vspace{2mm}
          \caption{ml100k.}
          \vspace{-2mm}
      \label{ml100k}
\begin{tabular}{lrrrr}
\toprule
Model &  HR@50 &  95\% CI &  MeanRanks &  95\% CI \\
\midrule
Random     &       0.047 &          0.030 &      476.3 &          39.5 \\
Popularity &       0.324 &          0.067 &      193.5 &          30.1 \\
UserKNN    &       0.553 &          0.071 &      107.3 &          23.8 \\
ItemKNN    &       0.569 &          0.071 &      102.9 &          22.0 \\
P3alpha    &       0.570 &          0.071 &      111.3 &          24.8 \\
RP3beta    &       0.571 &          0.071 &       95.4 &          19.8 \\
PureSVD    &       0.577 &          0.071 &      108.7 &          24.9 \\
\rowcolor[gray]{0.9}
MultiVAE   &       0.579 &          0.071 &       92.1 &          19.6 \\
GFCF       &       0.582 &          0.071 &       97.7 &          21.6 \\
\rowcolor[gray]{0.9}
MultiDAE   &       0.597 &          0.070 &       85.8 &          18.6 \\
ALS        &       0.598 &          0.070 &       96.7 &          22.0 \\
Ease       &       0.600 &          0.070 &       99.1 &          23.9 \\
SLIM       &       0.604 &          0.070 &      101.5 &          23.2 \\
\bottomrule
\end{tabular}
    \end{center}
  \end{subtable}
  \quad
  \begin{subtable}[t]{0.32\textwidth}
    \begin{center}
    \vspace{2mm}
          \caption{ml1m.}
          \vspace{-2mm}
      \label{ml1m}
\begin{tabular}{lrrrr}
\toprule
Model &  HR@50 &  95\% CI &  MeanRanks &  95\% CI \\
\midrule
Random     &       0.018 &          0.007 &     1518.1 &          50.0 \\
Popularity &       0.251 &          0.024 &      410.6 &          29.1 \\
ItemKNN    &       0.441 &          0.028 &      276.3 &          30.2 \\
UserKNN    &       0.449 &          0.028 &      258.4 &          29.0 \\
P3alpha    &       0.456 &          0.028 &      296.3 &          31.6 \\
RP3beta    &       0.462 &          0.028 &      242.1 &          26.3 \\
PureSVD    &       0.471 &          0.028 &      271.1 &          32.3 \\
GFCF       &       0.487 &          0.028 &      219.9 &          25.6 \\
\rowcolor[gray]{0.9}
MultiVAE   &       0.492 &          0.028 &      180.3 &          18.5 \\
SLIM       &       0.496 &          0.028 &      208.4 &          23.1 \\
\rowcolor[gray]{0.9}
MultiDAE   &       0.503 &          0.028 &      173.8 &          18.1 \\
ALS        &       0.504 &          0.028 &      216.8 &          25.6 \\
Ease       &       0.512 &          0.028 &      224.4 &          27.1 \\
\bottomrule
\end{tabular}
    \end{center}
  \end{subtable}
  \quad
  \begin{subtable}[t]{0.32\textwidth}
    \begin{center}
    \vspace{2mm}
          \caption{ml20m.}
          \vspace{-2mm}
      \label{ml20m}
\begin{tabular}{lrrrr}
\toprule
Model &  HR@50 &  95\% CI &  MeanRanks &  95\% CI \\
\midrule
Random     &       0.004 &          0.001 &     6814.6 &          46.6 \\
Popularity &       0.241 &          0.005 &      695.9 &          14.7 \\
P3alpha    &       0.432 &          0.006 &      957.8 &          24.9 \\
ItemKNN    &       0.462 &          0.006 &      559.9 &          18.4 \\
PureSVD    &       0.469 &          0.006 &      808.5 &          30.9 \\
RP3beta    &       0.476 &          0.006 &      374.7 &          12.6 \\
UserKNN    &       0.496 &          0.006 &      517.4 &          18.6 \\
GFCF       &       0.516 &          0.006 &      385.9 &          17.5 \\
ALS        &       0.519 &          0.006 &      494.7 &          22.2 \\
\rowcolor[gray]{0.9}
MultiVAE   &       0.522 &          0.006 &      253.8 &           7.8 \\
SLIM       &       0.523 &          0.006 &      405.1 &          14.8 \\
Ease       &       0.548 &          0.006 &      744.3 &          30.4 \\
\rowcolor[gray]{0.9}
MultiDAE   &       0.564 &          0.006 &      197.7 &           6.6 \\
\bottomrule
\end{tabular}
    \end{center}
  \end{subtable}
  \quad
  \begin{subtable}[t]{0.32\textwidth}
    \begin{center}
    \vspace{2mm}
          \caption{lastfm.}
          \vspace{-2mm}
      \label{lastfm}
\begin{tabular}{lrrrr}
\toprule
Model &  HR@50 &  95\% CI &  MeanRanks &  95\% CI \\
\midrule
Random     &       0.024 &          0.015 &     1374.2 &          82.2 \\
Popularity &       0.206 &          0.041 &      660.6 &          78.9 \\
PureSVD    &       0.464 &          0.051 &      337.8 &          64.7 \\
UserKNN    &       0.467 &          0.051 &      311.0 &          60.4 \\
P3alpha    &       0.491 &          0.051 &      220.0 &          44.0 \\
ALS        &       0.492 &          0.051 &      295.4 &          58.9 \\
ItemKNN    &       0.505 &          0.051 &      193.2 &          37.8 \\
RP3beta    &       0.508 &          0.051 &      191.5 &          38.1 \\
Ease       &       0.522 &          0.051 &      287.5 &          59.0 \\
GFCF       &       0.527 &          0.051 &      205.0 &          43.3 \\
SLIM       &       0.530 &          0.051 &      201.6 &          41.4 \\
\rowcolor[gray]{0.9}
MultiVAE   &       0.540 &          0.051 &      172.3 &          34.6 \\
\rowcolor[gray]{0.9}
MultiDAE   &       0.544 &          0.051 &      167.7 &          33.9 \\
\bottomrule
\end{tabular}
    \end{center}
  \end{subtable}
  \quad
  \begin{subtable}[t]{0.32\textwidth}
    \begin{center}
    \vspace{2mm}
          \caption{kuai.}
          \vspace{-2mm}
      \label{kuai}
\begin{tabular}{lrrrr}
\toprule
Model &  HR@50 &  95\% CI &  MeanRanks &  95\% CI \\
\midrule
Random     &       0.016 &          0.015 &     1419.9 &          98.9 \\
P3alpha    &       0.318 &          0.054 &      601.7 &          90.2 \\
\rowcolor[gray]{0.9}
MultiVAE   &       0.319 &          0.054 &      567.4 &          84.7 \\
Popularity &       0.319 &          0.054 &      588.9 &          86.0 \\
GFCF       &       0.320 &          0.054 &      590.4 &          87.6 \\
ItemKNN    &       0.322 &          0.055 &      621.6 &          90.8 \\
RP3beta    &       0.322 &          0.055 &      578.9 &          87.2 \\
\rowcolor[gray]{0.9}
MultiDAE   &       0.323 &          0.055 &      577.3 &          84.9 \\
UserKNN    &       0.323 &          0.055 &      598.6 &          87.7 \\
PureSVD    &       0.327 &          0.055 &      628.8 &          93.6 \\
ALS        &       0.330 &          0.055 &      599.8 &          90.7 \\
Ease       &       0.330 &          0.055 &      617.3 &          92.7 \\
SLIM       &       0.331 &          0.055 &      607.4 &          89.0 \\
\bottomrule
\end{tabular}
    \end{center}
  \end{subtable}
  \quad
  \begin{subtable}[t]{0.32\textwidth}
    \begin{center}
    \vspace{2mm}
          \caption{bookx.}
          \vspace{-2mm}
      \label{bookx}
\begin{tabular}{lrrrr}
\toprule
Model &  HR@50 &  95\% CI &  MeanRanks &  95\% CI \\
\midrule
Random     &       0.001 &          0.001 &    17270.8 &         372.2 \\
Popularity &       0.055 &          0.008 &     9118.9 &         395.9 \\
PureSVD    &       0.109 &          0.012 &    10701.4 &         450.7 \\
ALS        &       0.117 &          0.012 &     9704.5 &         430.6 \\
\rowcolor[gray]{0.9}
MultiVAE   &       0.129 &          0.012 &     4976.9 &         275.0 \\
GFCF       &       0.132 &          0.013 &     4852.8 &         282.9 \\
\rowcolor[gray]{0.9}
MultiDAE   &       0.137 &          0.013 &     4704.4 &         262.9 \\
UserKNN    &       0.137 &          0.013 &     7749.4 &         373.8 \\
ItemKNN    &       0.145 &          0.013 &    10098.6 &         400.6 \\
P3alpha    &       0.153 &          0.013 &     8949.8 &         392.8 \\
Ease       &       0.156 &          0.013 &    10142.6 &         463.3 \\
RP3beta    &       0.157 &          0.014 &     7526.9 &         374.7 \\
SLIM       &       0.161 &          0.014 &     8243.6 &         376.2 \\
\bottomrule
\end{tabular}
    \end{center}
  \end{subtable}
  \quad
  \begin{subtable}[t]{0.32\textwidth}
    \begin{center}
    \vspace{2mm}
          \caption{jester.}
          \vspace{-2mm}
      \label{jester}
\begin{tabular}{lrrrr}
\toprule
Model &  HR@50 &  95\% CI &  MeanRanks &  95\% CI \\
\midrule
Random     &       0.633 &          0.009 &       40.8 &           0.5 \\
PureSVD    &       0.837 &          0.007 &       23.6 &           0.5 \\
Popularity &       0.891 &          0.006 &       20.0 &           0.4 \\
ItemKNN    &       0.912 &          0.006 &       18.3 &           0.4 \\
P3alpha    &       0.927 &          0.005 &       17.4 &           0.4 \\
GFCF       &       0.928 &          0.005 &       17.1 &           0.4 \\
ALS        &       0.936 &          0.005 &       15.4 &           0.3 \\
Ease       &       0.939 &          0.005 &       14.9 &           0.3 \\
\rowcolor[gray]{0.9}
MultiVAE   &       0.939 &          0.005 &       15.7 &           0.3 \\
RP3beta    &       0.939 &          0.005 &       16.0 &           0.3 \\
SLIM       &       0.939 &          0.005 &       15.1 &           0.3 \\
UserKNN    &       0.940 &          0.005 &       15.4 &           0.3 \\
\rowcolor[gray]{0.9}
MultiDAE   &       0.947 &          0.004 &       14.7 &           0.3 \\
\bottomrule
\end{tabular}
    \end{center}
  \end{subtable}
  \quad
  \begin{subtable}[t]{0.32\textwidth}
    \begin{center}
    \vspace{2mm}
          \caption{amazon-e.}
          \vspace{-2mm}
      \label{amazon-e}
\begin{tabular}{lrrrr}
\toprule
Model &  HR@50 &  95\% CI &  MeanRanks &  95\% CI \\
\midrule
Random     &       0.001 &          0.000 &    22423.5 &         160.7 \\
Popularity &       0.063 &          0.003 &     9779.2 &         153.7 \\
PureSVD    &       0.070 &          0.003 &    12358.1 &         179.7 \\
ALS        &       0.089 &          0.004 &    12863.8 &         187.2 \\
GFCF       &       0.114 &          0.004 &    10004.1 &         165.5 \\
\rowcolor[gray]{0.9}
MultiVAE   &       0.114 &          0.004 &     7169.4 &         128.3 \\
UserKNN    &       0.119 &          0.004 &    14265.9 &         178.9 \\
\rowcolor[gray]{0.9}
MultiDAE   &       0.126 &          0.004 &     6424.7 &         120.2 \\
Ease       &       0.128 &          0.004 &    17552.1 &         221.5 \\
ItemKNN    &       0.128 &          0.004 &    15762.3 &         178.8 \\
P3alpha    &       0.129 &          0.004 &    14208.5 &         178.4 \\
RP3beta    &       0.129 &          0.004 &    14341.7 &         178.4 \\
SLIM       &       0.132 &          0.004 &    12844.7 &         174.0 \\
\bottomrule
\end{tabular}
    \end{center}
  \end{subtable}
  \quad
  \begin{subtable}[t]{0.32\textwidth}
    \begin{center}
    \vspace{2mm}
          \caption{netflix.}
          \vspace{-2mm}
      \label{netflix}
\begin{tabular}{lrrrr}
\toprule
Model &  HR@50 &  95\% CI &  MeanRanks &  95\% CI \\
\midrule
Random     &       0.003 &          0.000 &     8794.4 &          32.7 \\
Popularity &       0.172 &          0.002 &     1005.6 &          12.2 \\
P3alpha    &       0.361 &          0.003 &     1759.7 &          25.7 \\
ItemKNN    &       0.374 &          0.003 &     1426.6 &          23.0 \\
UserKNN    &       0.387 &          0.003 &     1144.4 &          20.4 \\
RP3beta    &       0.389 &          0.003 &     1415.7 &          23.0 \\
PureSVD    &       0.398 &          0.003 &     1281.5 &          24.2 \\
GFCF       &       0.404 &          0.003 &      556.6 &          12.3 \\
\rowcolor[gray]{0.9}
MultiVAE   &       0.417 &          0.003 &      420.9 &           6.8 \\
ALS        &       0.424 &          0.003 &      742.4 &          16.5 \\
SLIM       &       0.450 &          0.003 &      830.3 &          17.2 \\
\rowcolor[gray]{0.9}
MultiDAE   &       0.462 &          0.003 &      327.8 &           5.8 \\
Ease       &       0.467 &          0.003 &     1099.6 &          23.0 \\
\bottomrule
\end{tabular}
    \end{center}
  \end{subtable}
  \vspace{1mm}
  \caption{Performance of all recommendation models over nine real-world datasets. The results on each dataset are ranked in an ascending order from top to bottom in terms of HitRate@50 (denoted as HR@50). "95\% CI" represents the 95\% confidence interval. Neural models are marked in grey.}
  \vspace{-8mm}
  \label{all_gen_results}
\end{table*}

\renewcommand{\arraystretch}{1}

 \textbf{Active Users vs. Inactive Users.}
We also analyze the performance differences between active and inactive users. Here active refers to users with the largest numbers of ratings, while inactive denotes the opposite (thresholds were picked based on percentiles). Our findings are as follows. First, in terms of active user subgroups, both neural models exhibit better performance for MeanRanks than for HitRate. However, they do not consistently outperform non-neural models. For example, SLIM is a much simpler non-neural model but achieves better performance compared to neural models in terms of both utility metrics. Second, when considering inactive users, MultiDAE achieves better performance, and the performance of most models remains stable for HitRate. However, the MeanRanks performance of most non-neural models declines for inactive users, while both neural models fare best. 
These observations imply that neural models demonstrate better robustness compared with non-neural models (especially on MeanRanks) between users with different activity levels.

 \textbf{Head Items vs. Tail Items.}
We finally compare the performance of the recommendation models on head and tail items. Again, head items refer to items with the largest number of implicit user feedback, while tail items are the opposite. The performance of each model is reported over two user subgroups partitioned by which type of item is held out randomly.
We observe that neural models do not outperform non-neural ones in terms of HitRate, especially for users whose hold-out items are head items. However, neural models do have an advantage when considering MeanRanks in both settings. Second, both neural models maintain their performance advantage over other non-neural models on tail items, while the performance of most other non-neural models drops on tail items.
Several non-neural models (e.g., ALS and PureSVD) show better performance on tail items, however, they are  behind  the neural models.
These observations suggest that neural models not only achieve better MeanRanks scores in general but also exhibit better robustness across head and tail items.

Based on the discussion above, our results indicate that neural models tend to have stronger robustness over warm/cold-start instances compared to non-neural ones. As a consequence, we suggest that neural models should be preferred if better MeanRanks or robustness is desired in practice.

\section{Related Work}

Researchers have long been noticing that the results of multiple recommender models are hard to compare due to differences in baseline implementations~\cite{said2014comparative}, dataset pre-processing strategies~\cite{konstan2013toward}, and evaluation methodologies~\cite{ekstrand2011rethinking}.
These findings have cast doubt on whether the reported performance increases of each new model can actually add up.
This leads researchers to re-investigate existing recommendation models~\cite{ferrari2019we,anelli2022top}. Most of the replication studies found it hard to obtain similar results as published previously~\cite{cremonesi2021progress}. This further boasts the claim that reproducibility and replicability in recommender systems are  central issues\cite{ferrari2019we,ludewig2019performance,rendle2019difficulty,armstrong2009improvements,anelli2022top}.
For example, Sedhain et al.~\cite{sedhain2016effectiveness} pointed out that simple linear classification models compare quite favorably with state-of-the-art; P{\'e}rez et al.~\cite{perez2022evaluation} found several neural recommendation models are to be non-reproducible; Kolesnikov et al.~\cite{kolesnikov2021ttrs} pointed out potential flaws with popular evaluation strategies.

 \textbf{Neural vs. Non-Neural Recommendation Models.}
Compared with non-neural models, neural models have been claimed to outperform traditional  recommender systems in recent years.
To the best of our knowledge, there are only three other relevant studies on the difference between neural and non-neural models. Anelli et al.~\cite{anelli2022top} found that neural models cannot outperform most traditional models, e.g., linear models and matrix factorization. Ludewig et al. ~\cite{ludewig2019performance} found similar results for session-based recommendation. Finally, Zhao et al.~\cite{zhao2022revisiting} argued that the training efficiency of neural models is significantly lower than that of other traditional models, such as linear models. This casts further doubt on whether or not it is worthwhile to adopt neural models in practice.

 \textbf{Recommender System Evaluation.}
It is worth noting that recommender systems do not exist in isolation but are part of a user-facing system~\cite{cremonesi2013user}. In real-world live recommender systems, any potential flaw can have a potentially huge negative impact on users.
Hence it is critical to systematically understand the properties of recommendation models offline before users are exposed to them.
Multiple studies have tried to systematically evaluate existing recommendation models~\cite{ferrari2019we,ludewig2019performance,rendle2019difficulty,armstrong2009improvements,anelli2022top}. For example, Sun et al.~\cite{sun2020we} argued that certain essential factors (e.g., data splitting methods and hyper-parameter tuning strategies) can influence the performance recommendation model dramatically. Therefore, they proposed standardized procedures for a more rigorous evaluation.
Other recent studies have also looked at other parts  of the whole evaluation pipeline~\cite{anelli2021elliot,jin2021towards,ludewig2019performance,ferrari2019we,latifi2022sequential,anelli2021reenvisioning}.
Furthermore, there are also existing studies focusing on specific factors in evaluation. 
(1) \textbf{Dataset Construction.} Datasets have been playing a critical role in delivering accurate recommendations in existing recommendation models. The strategies to pre-process and construct the dataset have attracted much research attention~\cite{meng2020exploring,li2020sampling,zhao2022revisiting,latifi2022sequential,sachdeva2022sampling}. For example, Sachdeva et al.~\cite{sachdeva2022sampling} evaluated the efficacy of sixteen different sampling strategies for benchmarking recommendation models.
(2) \textbf{Baselines.} Whether appropriate baselines are included or not is also a critical factor in the evaluation of recommender systems. As an example, Ji et al.~\cite{ji2020re} proposed to re-visit the commonly chosen baselines in recommender systems.
(3) \textbf{Model Optimization.} Model optimization can also exert a significant influence on the performance of recommender systems, e.g., hyper-parameter tuning strategies. For example, Zhao et al.~\cite{zhao2022revisiting} argued that the search range of hyper-parameters often affects the performance substantially, while Anelli et al.~\cite{anelli2019discriminative} pointed out that only tuning a few parameters will usually help achieve adequate performance.
(4) \textbf{Evaluation Metrics.} It is also important to study whether the adopted evaluation metrics are appropriate or not~\cite{krauth2020offline,zhao2022revisiting,felicioni2021measuring,dacrema2022offline,latifi2022sequential}. For instance, Zhao et al.~\cite{zhao2022revisiting} found that adopting sampling-based metrics could introduce bias in the evaluation process.

 \textbf{Significance of Our Work.}
Despite the progress in evaluating recommendation models, there is not much work comparing different types of recommendation models~\cite{jin2021towards,ludewig2019performance}. However, we note that it is a critical issue since such an understanding facilitates the researchers and engineers to choose appropriate models under different application scenarios.
Meanwhile, existing insights on the difference between neural and non-neural models are also limited. For example, Anelli et al.~\cite{anelli2022top} conducted an evaluation between neural and non-neural models. However, they adopted weak generalization as the evaluation strategy, which we argue is not representative of real-world recommendation systems that are needed to make recommendations to new users. Ludewig et al.~\cite{ludewig2019performance} compared the performance between neural and non-neural models. However, their work only focuses on session-based recommendation, ignoring the most common recommendation problem with implicit user feedback. Zhao et al.~\cite{zhao2022revisiting} evaluated popular recommendation models considering different factors. However, they did not provide a comprehensive discussion of the performance between these two types of recommendation models.
Different from most existing work on recommendation model evaluation, this paper provides comprehensive evaluation results under a strong generalization setting, which gives a better understanding of the characteristics of the two types of recommendation models.

\section{Conclusion and Discussion}

In this work, we present a thorough investigation into the performance differences between neural recommendation models and non-neural ones. We empirically explored what type of benefit one may gain when using neural models instead of traditional non-neural ones for recommendation in practice. We introduced a number of practical and diverse evaluation strategies for benchmarking and then conducted extensive experiments on nine publicly available real-world datasets over 13 popular recommendation models. 
We show that in most cases, neural models do not show superior performance over other non-neural models when considering  HitRate. They do, however, fare better in terms of MeanRanks. In addition, neural models, on average, showed a higher level of diversity, were better at capturing item-item semantics, and exhibited stronger robustness in warm-start and cold-start scenarios.

To the best of our knowledge, this is the first work to investigate the performance differences between neural and non-neural recommendation models in top-$n$ recommendation with implicit user feedback. 
Investigations in terms of the widely studied memorization, under-explored generalization, and rarely discussed subgroup-specific performance are included.
Future works based on this paper may explore whether neural models exhibit superiority over other popular metrics and whether the benefits of neural models are worth the training costs. However, we note that they are beyond the scope of this paper.
We hope this work can help practitioners with their choice of recommendation models, inspire more research around understanding recommender models, and facilitate better model design in the future.

\section{Acknowledgements}
Yushun Dong and Jundong Li are supported by the National Science Foundation under grants IIS-2006844, IIS-2144209, and IIS-2223769.

\bibliographystyle{ACM-Reference-Format}
\bibliography{references}

\end{document}